\documentclass[journal]{IEEEtran}

\usepackage{soul}
\IEEEoverridecommandlockouts
\usepackage{float}
\usepackage{url}

\usepackage[linesnumbered,ruled,vlined]{algorithm2e}
\usepackage{setspace}
\usepackage{cite}
\usepackage[compatibility=false]{caption}
\usepackage{subcaption}
\usepackage{graphicx}  
\usepackage{tikz} 
\usepackage{algorithmic}
\usepackage{romannum}
\usepackage{pdfpages}
\usepackage{psfrag}
\usepackage{textcomp}
\usepackage{amsmath,amssymb,amsfonts}
\usepackage[utf8]{inputenc}
\usepackage{enumitem}
\usepackage{multicol}
\usepackage{lipsum} 
\usepackage{siunitx}
\usepackage{footmisc}
\usepackage{xcolor}
\usepackage{pgfplots}
  \pgfplotsset{compat=newest}
  \usetikzlibrary{plotmarks}
  \usetikzlibrary{arrows.meta}
  \usepgfplotslibrary{patchplots}
  \usepackage{grffile}
\begin{document}
\title{Radio over Fiber with Cascaded Structure: Algorithm for Uplink Positioning
}
\author{Dexin Kong,~\IEEEmembership{Graduate Student Member, IEEE,}
        Diana Pamela Moya Osorio,~\IEEEmembership{Senior Member,~IEEE,}
        and~Erik G. Larsson,~\IEEEmembership{Fellow,~IEEE}
  \thanks{The authors are with the Dept. of Electrical Engineering (ISY), Link\"oping University, 581 83 Link\"oping, Sweden (email: dexin.kong@liu.se; diana.moya.osorio@liu.se; erik.g.larsson@liu.se).}
  \thanks{This work is an extended version of our conference paper presented at IEEE SPAWC 2024 \cite{MyConference}. This work was supported in part by ELLIIT, and in part by the 6G Tandem project funded by the European Union’s Horizon Europe research and innovation programme under Grant Agreement No 101096302.}}
\maketitle 

\begin{abstract}
Recent advancements in polymer microwave fiber (PMF) technology have created significant opportunities for robust, low-cost, and high-speed sub-terahertz (THz) radio-over-fiber communications. Recognizing these potential benefits, this paper explores a novel radio-over-fiber (RoF) structure that interconnects multiple radio units (RUs), booster units (BUs), and a central unit (CU) in cascade via fiber, envisioning its application in indoor scenarios. This structure creates a number of research opportunities when considering cascaded distortion effects introduced by non-linear power amplifiers (PAs) and the propagation channel over the fiber. 

We propose maximum-likelihood and non-linear least-squares algorithms to estimate the entry RU and the time-of-arrival between the RoF and the user equipment, where estimating the entry RU is equivalent to estimating the propagation distance along the RoF stripe. For the case of linear PAs, we derive the Cramér-Rao lower bound to benchmark the performance of the estimators. Finally, we investigate the use of the system for uplink positioning. Our simulation results demonstrate that the proposed estimators perform satisfactorily even with the cascaded effects of non-linear PAs, and that the deployment of this RoF structure can enable new cost-effective opportunities for high-resolution positioning in indoor scenarios.
In the numerical evaluation, we also use measured PMF characteristics for high-density polyethylene fibers. 
\end{abstract}
 \vspace{0.2cm}
\begin{IEEEkeywords}
Non-linear power amplifiers, polymer microwave fiber, radio over fiber, high-resolution positioning
\end{IEEEkeywords}
\section{Introduction}
\IEEEPARstart{N}EXT-generation wireless communication systems are expected to support ubiquitous wireless connectivity. 
To achieve this goal, one crucial limitation is the spectrum resources. Although sixth-generation (6G) communication systems are expected to adopt frequency bands that strike a balance between capacity and coverage, particularly within the new mid-band spectrum (7.125 to 24.25 GHz)~\cite{bazzi2025upper}, the sub-terahertz (sub-THz) range (90 to 300 GHz) is anticipated to enable high-precision sensing and short-range positioning applications~\cite{THzsurvey,cai6G,6Gwhitepaper}. Consequently, research on sub-THz technologies is expected to continue advancing~\cite{shafi2025industrial}. In this work, we present a novel system operating in the sub-THz band to support ultra-high data rates, low latency, and enhanced positioning accuracy. This is largely due to the wide bandwidth available in the sub-THz spectrum, which enables data transmission rates of several terabits per second and provides high spatial resolution.
 
Although there has been abundant research on the performance of sub-THz wireless communication systems~\cite{THzdata,THzXR,SubTHzPosition}, their implementation is still challenging~\cite{THzimplementationchallenge}. Considering the cost of sub-THz radio hardware and the limited coverage~\cite{bjornson2019massive}, a low-cost and densely deployable solution is desirable. Recent developments in polymer microwave fiber (PMF) have motivated the development of radio-over-fiber (RoF) communications, which offer great opportunities for low-cost implementations of sub-THz wireless systems~\cite{analysisHE11,Foam-cladded,polymerblend}. RoF systems can be implemented in different ways. Herein, we consider a system with a cascaded structure, where multiple radio units (RUs) and booster units (BUs) are interconnected via low-cost PMFs and finally connected to a central unit (CU). Each RU is equipped with antennas, a power amplifier (PA), and other radio frequency (RF) components; see Section~\ref{System overview} for details. The only difference between an RU and a BU is the absence of antennas in a BU. The RU capturing or transmitting the sub-THz signals acts as an access point, and all other units between the entry/exit RU and the CU only amplify and forward the signals. All signal processing is performed at the CU. Since there are no digital signal processing units in the RUs and BUs, the system is significantly less costly than a dense deployment of radio transceivers.

\begin{figure}[t]
    \centering
    \includegraphics[width=1\linewidth]{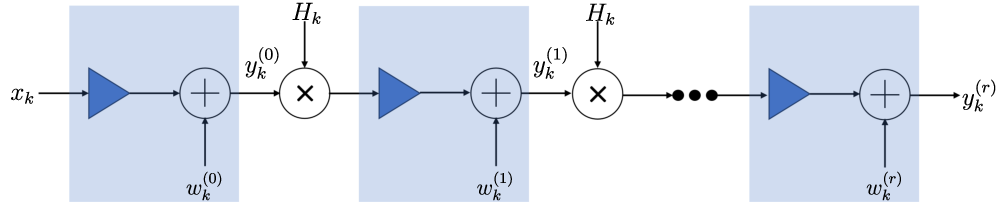}
    \caption{Signal model for the RoF system with PAs and PMFs connected in cascade.}
    \label{signal model linear}
\end{figure}
Fig.~\ref{signal model linear} illustrates the cascade RoF system under consideration (the different variables will be introduced later). In this system, the signal propagates over a short-range sub-THz wireless channel and over multiple dispersive fibers within the RoF system. The salient feature of this system is that the propagation distance over the sub-THz wireless channel is only a small fraction of the distance between the CU and the user equipment (UE). According to measurements reported in \cite{Frida}, state-of-the-art PMFs attenuate sub-THz signals less than 5\,dB per meter in the D-band (110 to 170\,GHz), while the free-space path loss in the D-band is around 70\,dB on the first meter and then follows an exponential law~\cite{cai6G}. Consequently, the overall propagation attenuation is significantly less in an RoF system compared to in a pure wireless sub-THz system with the same coverage.

\begin{figure}[t]
    \centering
\includegraphics[width=1\linewidth]{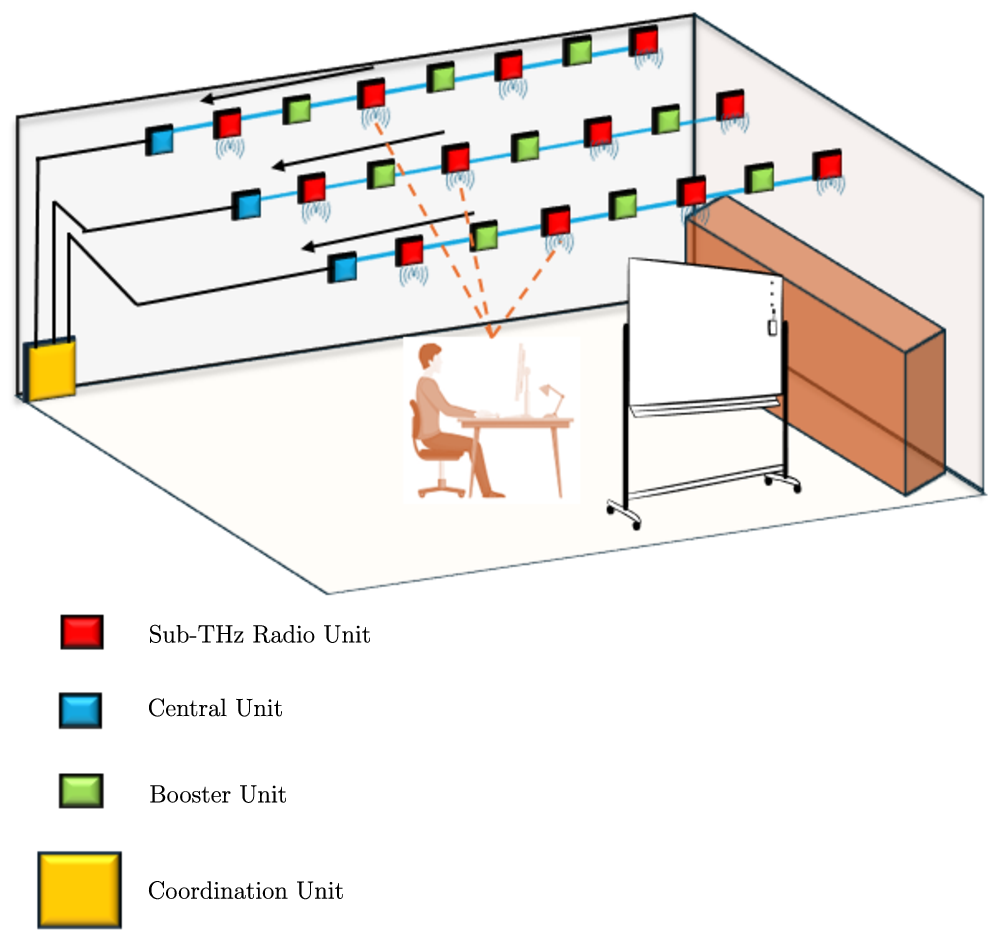}
    \caption{An RoF system deployment in an indoor scenario.}
    \label{indoor scenario}
\end{figure}

The power amplifiers (PAs) are identified as the dominant source of power consumption. The energy efficiency (EE) of PAs is improved if a small back-off is adopted, i.e., the PAs are allowed to work in the non-linear regime~\cite{6Gwaveform}. Therefore, at each stage of the RoF system, the signals are distorted by a non-linear (PA) and a segment of dispersive PMF, with the latter being considered a linear time-invariant (LTI) system. After multiple stages, the signal will be subjected to a cumulative distortion, which we exploit to estimate the propagation distance over the RoF stripe. In this paper, each UE is assumed to transmit data to three RoF stripes that are deployed in parallel and mounted at the ceiling, as shown in Fig~\ref{indoor scenario}. The CU at each stripe jointly estimates the propagation distance over the stripe, as well as the time-of-arrival (ToA) with a clock bias between the UE and the stripe. Based on estimates of ToA and propagation distance from three CUs, trilateration is used to locate the UE. A detailed geometry model and problem formulation will be presented in Section~\ref{PF and SM}. 

\textbf{Contributions:} The specific contributions of this paper are 
(i) an analysis of the effects of non-linear PAs and dispersive PMFs connected in a cascade, and an analysis of the uplink (UL) signal; (ii) a maximum-likelihood (ML)  framework and a non-linear least-squares (NLS) framework tailored to the cascade-structured RoF for estimating the entry RU and the ToA with bias; (iii) a thorough evaluation on the performance of the estimators with PAs working in the linear regime via a  Cramér-Rao lower bound (CRLB) analysis; (iv) numerical results (using measured PMF dispersion characteristics and different values of non-linear factors) demonstrate that the proposed estimators perform well on the propagation distance estimation even in the presence of the effects of cascaded non-linear PAs; and (v) we demonstrate the usability of the proposed system for high-resolution positioning in indoor scenarios. To the best of our knowledge, this is the first study of an RoF system with segments of dispersive PMFs and with non-linear amplifiers connected in cascade. The proposed system and analysis have the potential to open up several novel research directions and opportunities for positioning and localization systems.

The rest of this paper is organized as follows. Section~\ref{Related Work} presents related works. Section~\ref{System overview} discusses the system overview and the structure of every component within an RoF stripe. Section~\ref{PF and SM} formulates the problem and the corresponding signal models. Section~\ref{Cramer-Rao Lower bound} evaluates the ML estimator in the linear regime by the CRLB. Section~\ref{simulation results} shows the applied channel measurement data and numerical results. Section~\ref{Use case} applies the proposed algorithm in an indoor positioning use case. Section~\ref{conclusion} concludes the paper.

\textit{Notations:} Boldface letters denote column vectors; regular letters denote scalars. $(\cdot)^\mathrm{H}$ denotes Hermitian transpose operation and $(\cdot)^\mathrm{T}$ denotes the transpose operation; $\mathbb{C}$ is the set of complex numbers; The $k$th element of a column vector $\boldsymbol{x}$ is denoted by the subscript as $\boldsymbol{x}_k$; The superscript $n$ of a vector/scalar $(\cdot)^{(n)}$ denotes the vector/scalar at the $n$th unit of an RoF stripe; $\lvert \cdot \rvert$ denotes the absolute value; $\lvert\lvert.\rvert \rvert$ denotes the $L^2$ norm; $\delta(\cdot)$ is the delta function; $\mathrm{det}(\cdot)$ calculates the determinant of a matrix; $\mathrm{diag}\{\cdots\}$ represents a diagonal matrix; $\mathbf{1}_N$ denotes a all one column vector with $N$ elements; Finally, $j=\sqrt{-1}$ is the imaginary unit.  
  
\section{Related work} \label{Related Work}
It is worth noting that this study considers an RoF system formed with only RF components and fibers. This implementation is novel and has not been investigated before, and it should not be confused with other RoF approaches that we introduce in the following.

In contrast to the system we consider, a traditional RoF system merges RF and optical fiber technologies, which typically contains lasers, an optical modulator/demodulator, optical amplifiers, optical fibers, a photomixer, and RF antennas. The mobile signal is carried by the modulated optical carriers and conveyed to the photomixer via a fiber. The photomixer then converts the optical carriers into RF electrical signals before the signals are transmitted by the RF antennas. Reference \cite{Millimeter_ROF_Survey} discusses traditional RoF systems in more detail. It is broadly acknowledged that traditional RoF systems present low attenuation and high cost-efficiency due to the use of low-cost fibers whose development is reported in~\cite{analysisHE11,Foam-cladded,polymerblend}. Moreover, since the signals in such RoF systems are well confined within the fiber, these systems are not restricted in bandwidth. The work in~\cite{THzRoFExperiment} experimentally demonstrates the RoF systems in the THz frequency band, which implements a fiber-wireless seamless network in an indoor scenario. The reported results demonstrate good performance in terms of carrier-to-noise ratio. However, this type of RoF system often suffers from severe non-linearity in optical components, such as lasers~\cite{laserNonlinearity}, and materials\cite{MaterialNonlinearity}, which represent the main challenge for RoF communications. 

It is also worth differentiating the proposed system from radio stripes (RS), which is a commonly envisioned implementation of distributed massive MIMO (D-MMIMO). D-MMIMO distributes the antennas in a large area to provide ubiquitous service for every UE. The work in~\cite{6GCFMIMO} provides a comprehensive overview of D-MMIMO in 6G. Although the D-MMIMO offers consistently high-quality service to all UEs, accurate synchronization between APs and the CU is challenging. On the other hand, RSs connect APs sequentially in one cable/fiber to the CU~\cite{OverviewRS}, which achieves simplified synchronization between APs via the shared bus.
The authors in~\cite{ULPositioningRS} investigate the high-accuracy positioning capability with RSs at the sub-6\,GHz band. However, previous works about RS consider that every AP is equipped with full signal processing abilities, which are considerably expensive at THz frequencies. In contrast, the unique structure considered in this paper allows dense and flexible deployment because the CU allocates all the resources and processes the signals. The studied system distinguishes itself from the distributed massive MIMO in the literature due to the absence of multiple antenna signal processing.
   
\section{System Overview} \label{System overview}
Consider the system illustrated in Fig.~\ref{indoor scenario}, where multiple RoF stripes are deployed in parallel to serve UEs in an indoor scenario. Each RoF stripe consists of multiple RUs, BUs, and one CU at the end. One coordination unit connects the RoF stripes with the servers and coordinates different RoF stripes. All RoF stripes are installed on the ceiling at the same height. The configurations of RU, BU, CU, and PMF will be discussed below. All other components are considered to be LTI systems. In this study, we assume that those components introduce attenuation but no significant dispersion.

\begin{figure}[!t]
\centering
\includegraphics[width=1.03\linewidth]{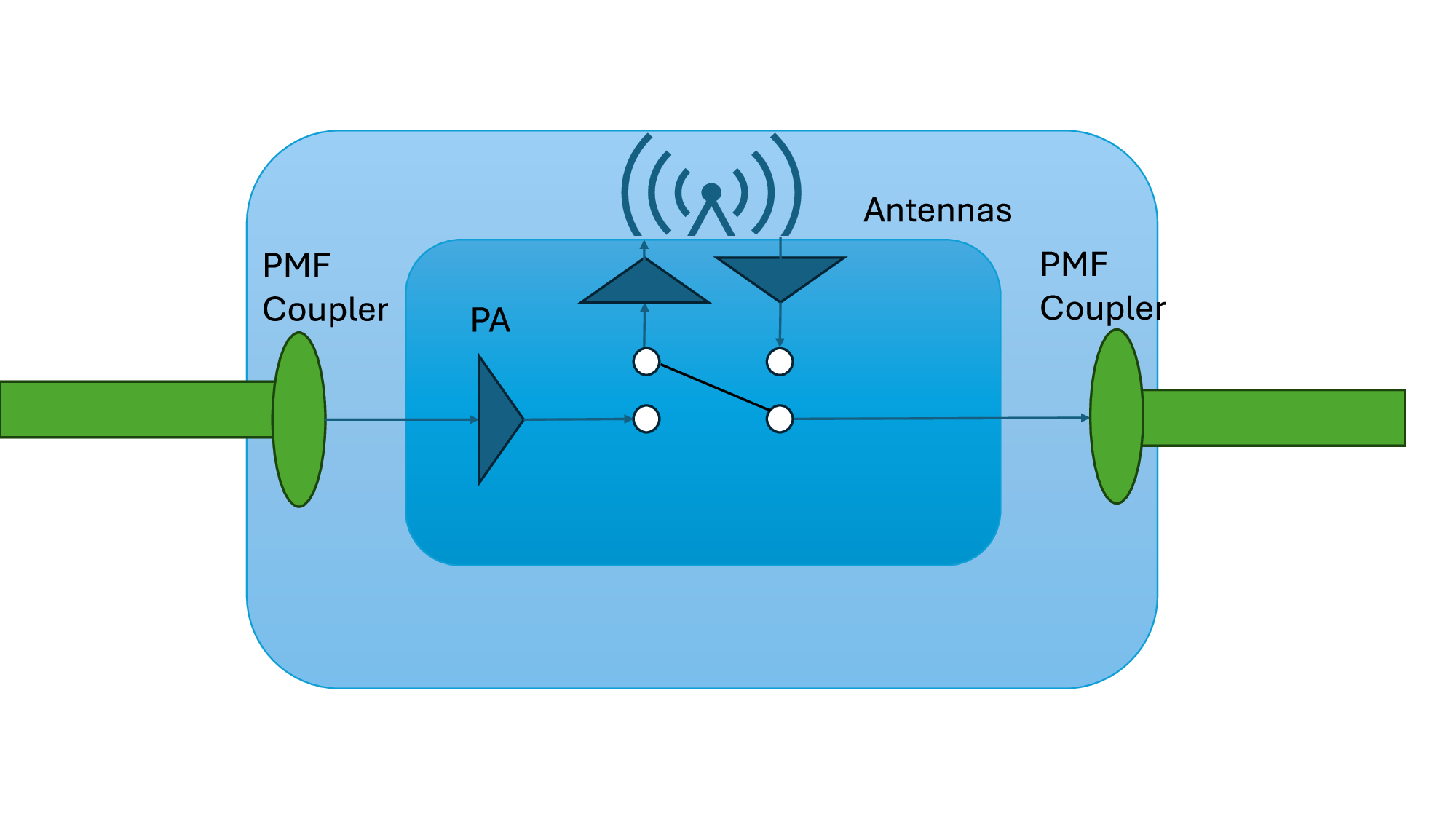}
    \caption{The structure of an RU.}
    \label{RU1}
\end{figure}

\begin{figure}[!t]
    \centering
\includegraphics[width=1.03\linewidth]{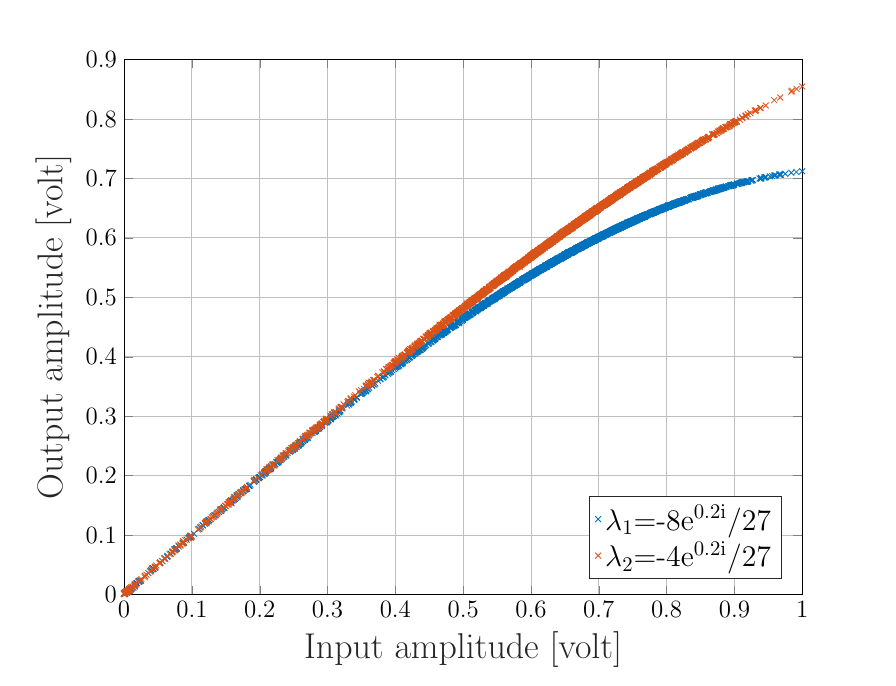}
    \caption{Characteristics of the PA with different values of the non-linearity factor.}
    \label{PA charact}
\end{figure}

\begin{figure}[!t]
 \centering\includegraphics[width=1\linewidth]{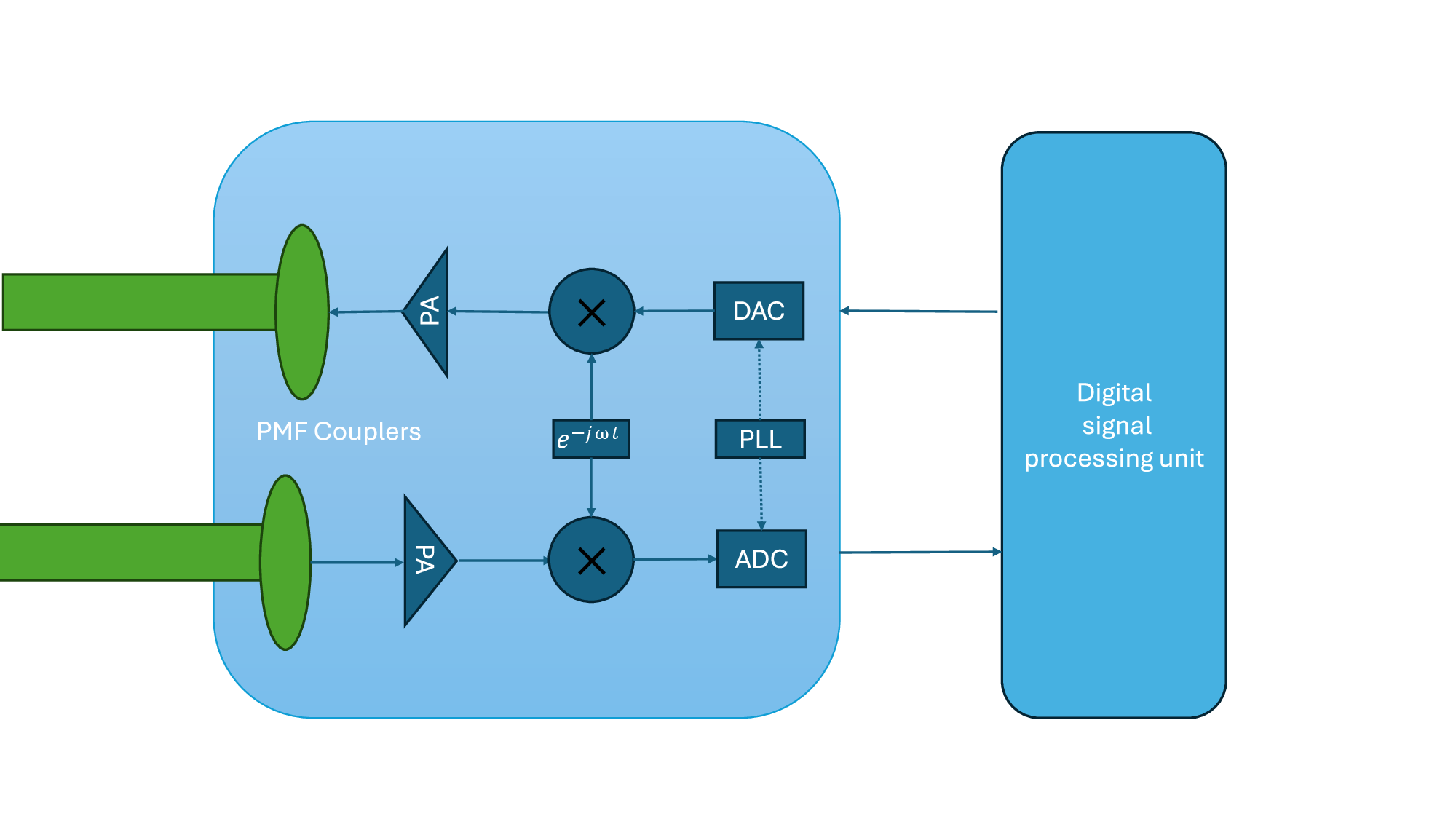}
    \caption{The structure of a CU.}
    \label{CU}
\end{figure}

\subsection{Polymer microwave fiber}
For a unit-length fiber segment, the channel impulse response with $L$ taps and amplitudes ($[\beta_0, \beta_1, \cdots, \beta_{L-1}]^\mathrm{T}$)\footnote{This model is valid in a non-linear system if the signals are properly oversampled~\cite{Thomas_ITC_2022}. We consider a third-order polynomial non-linear distortion; hence, the signals must be oversampled at least three times.} can be expressed as
\begin{equation}
    h_{n}=\sum_{l=0}^{L-1} \beta_l \delta(n-l), \label{eq: th}
\end{equation}
where $n$ is the time index. Then, a fiber with $r$ unit length is equivalent to connecting $r$ LTI systems with the same impulse response $h_n$ in a cascade, whose impulse response is the $r$ times convolution of $h_n$ and is expressed as
\begin{equation}
    \Tilde{h}_{n}=\underbrace{h_{n} * h_{n}*\cdots*h_{n}}_{r\,\text{convolutions}},\label{eq: hn}
\end{equation}
where all factors in \eqref{eq: hn} are modeled as in~\eqref{eq: th}. In the frequency domain, we consider the frequency response for a set of $K$ discrete frequencies, $\boldsymbol{f}=[f_0,\cdots,f_{K-1}]^\mathrm{T}$. Thus, the frequency domain representation of \eqref{eq: hn} is given by the multiplication of frequency responses. Let $\boldsymbol{H}$$=$$[H_0, \cdots, H_{K-1}]^\mathrm{T}$ represent the frequency response of a unit-length fiber. Then, the frequency response of a unit-length fiber at the $k$th frequency $H_k$ is the Fourier transform of its impulse response given by
\begin{equation}
        H_k= \sum_{n=0}^{N-1} h_n e^{-j2\pi f_k n T_s},
\end{equation}
where $T_s$ is the sampling time interval. Hence, the frequency response of an $r$-units-long fiber, at the $k$th frequency, can be expressed as 
\begin{equation}
    \Tilde{H}_k= \underbrace{H_k H_k \cdots H_k}_{r\,\text{multiplications}} = H_k^r.\label{eq: H}
\end{equation} 

\subsection{Radio Unit \& Booster Unit}
RUs are responsible for capturing incoming sub-THz signals, amplifying them, and forwarding them to the next component in the same stripe. Each RU consists of PMF couplers, PAs, antennas, and a switch as illustrated in Fig.~\ref{RU1}. For the operation of the RoF system, we consider that each RU is equipped with a switch that selectively receives signals, either from the PMF or the sub-THz antennas. When the RU detects signals from the antennas, the switch disconnects the PMF input, allowing only the signals captured by the antennas to pass through. Therefore, each RU only admits signals either from the antennas or from the PMF. If there are two RUs in a stripe capturing signals, only the RU that is closer to the CU would act as an access point.

In case the RU structure does not implement a switch, thus allowing to receive signals coming from both sides, antennas and the PMF, a superposition of the received signals from different RUs would be received at the CU. This case is not addressed in the present work, as its treatment requires an entirely separate analysis. 

BUs are deployed between any two adjacent RUs in an RoF stripe, as shown in Fig.~\ref{indoor scenario}. As previously mentioned, the BUs have the same structure as the RUs without the presence of antennas. The BUs amplify the signals and forward them to the next component.

To understand the impact of PAs, we examine both regimes of operation of PAs, linear and non-linear. The effects of non-linear PAs are characterized with a memoryless polynomial model. We use a third-order memoryless polynomial model with a factor $\lambda$, as higher-order terms are significantly less than the third-order term  \cite{nonlinearPA,majidi2014analysis}. A real-valued non-linear factor indicates only amplitude distortion, while a complex-valued non-linear factor indicates both phase and amplitude distortion. In practice, for a specific PA, $\lambda$ can be measured in a laboratory. The relationship between the input $x_n$ and output $y_n$ of a PA is given by\begin{itemize}
    \item Linear regime: $y_n=G x_n$,
    \item Non-linear regime: $y_n=G(x_n+\lambda x_n \lvert x_n \rvert^2),$
\end{itemize} 
where $G$ is the amplification coefficient. Fig.~\ref{PA charact} illustrates the characteristics of PA with different values of non-linear factors, where we can observe that the non-linearity arises with a signal magnitude of $\sim 0.4$ volts.

Due to the absence of a digital signal processing unit at a RU/BU, every RU/BU is incapable of processing signals. This design lowers the price of each RU/BU, but also disables the RU/BU to mitigate signal impairments introduced by PAs and PMFs. As a consequence, this results in the accumulation of signal distortion along the propagation in an RoF system.

\subsection{Central unit}
The CU is in charge of allocating resources, synchronizing the units in a stripe, and processing data. Each CU comprises PAs, mixers, couplers, ADC/DAC, phase-locked loops (PLL), and a digital signal processing unit (see Fig.~\ref{CU}). The CU is capable to operate in full-duplex mode to communicate with one UE simultaneously. 

\section{Problem formulation and signal model}\label{PF and SM}
\begin{figure}[t]
        \centering
        \includegraphics[width=1.03\linewidth]{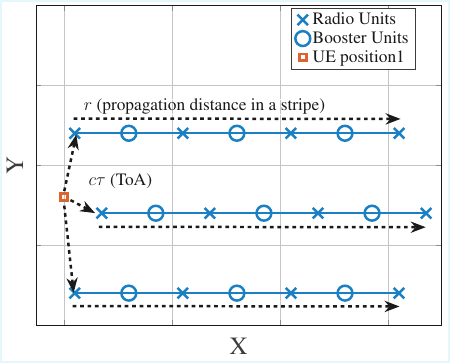}
        \caption{Positioning of the UE with three RoF stripes.}
        \label{fig: positioning_scenario}
    \end{figure}

To localize a UE in a two-dimensional space, at a minimum, three observations are required. With the system structure of the cascaded-structured RoF, three stripes are required to receive the signals from a UE simultaneously. As shown in Fig.~\ref{fig: positioning_scenario}, each stripe jointly estimates the propagation distance (denoted as $r$) in the stripe as well as the ToA (denoted as $\tau$) of the incoming signals. The estimated propagation distances are used to locate three RUs that serve the UE, and the estimated ToAs are used for trilateration.

Before deriving the received signal model, we emphasize the key assumptions of the paper. The RoF stripes are assumed to be synchronized in time and phase via wired connections. From the perspective of practical operation, it is challenging to synchronize the UE and RoF stripes. Therefore, it is assumed that the UE has a clock bias and phase offset relative to all RoF stripes, which are unknown to the CU. The sub-THz signals propagate both over the air and over the fiber. Due to the sparse nature of the sub-THz wireless channel, the line-of-sight (LoS) components are considered to be dominant, whose path loss and ToA are not available to the CU. On the other hand, a fixed-length fiber segment is an LTI system whose impulse response can be measured in a laboratory. According to \eqref{eq: th} to \eqref{eq: H}, the channel information of a fiber of any length is available to the CU. Similarly, the PA characteristics ($G, \lambda$, and noise variance at room temperature $\sigma^2$) are stable in the room temperature and can be measured after manufacture. In practice, a small deviation of PA characteristics does not have a significant effect, which is discussed in Section~\ref{sec: non-ideal PA}. In conclusion, the unknown parameters are i) $A$,  a complex value absorbing the effects of the wireless channel and the phase offset between the UE and the RoF; ii) $\tau$,  the propagation delay including ToA and a clock bias between the UE and the stripe; iii) $r$, the propagation distance in a stripe; iv) the coordinates of the UE. In particular, each CU/stripe estimates one propagation delay and propagation distance in the stripe, which are then used for trilateration to obtain the coordinates of the UE. All other variables in the signal model, as shown in the sections below, are known.

The RUs and BUs are uniformly distributed, which means that the distance between two consecutive units on an RoF stripe is fixed, and the distance between two adjacent RoF stripes is also fixed. Note that the stripes are shifted slightly relative to one another to ensure that the UE is not served by three RUs that lie on a line. This ensures that the geometric dilution of precision (GDOP) is favorable (see calculation in Section~\ref{Use case}). The coordinate of the $i$th position of UE is given by $(U_x^{(i)}, U_y^{(i)}, U_z)$, and the location of the CU in the $m$th stripe is given by $(P_x^{(m)},P_y^{(m)},0)$. Therefore, the distance between the UE and the RU in service at the $m$th stripe is  
\begin{equation}
    d_{m,i}=\sqrt{\lvert U_x^{(i)}-(P_x^{(m)}-r)\rvert^2 + \lvert U_y^{(i)}-P_y^{(m)}\rvert^2 +U_z^2}. \label{eq: wireless distance}
\end{equation}
The distance between the UE and the ceiling, $U_z$, is assumed to be fixed and known to the CU. The proposed positioning algorithm is based on ToA with bias, whose GDOP of a certain UE position $(U_x^{(i)}, U_y^{(i)})$ can be calculated according to the standard formulas~\cite{Shin_GDOP_2002}. The GDOP is defined as 
\begin{equation}
    GDOP(i)=\sqrt{\mathrm{tr}\left(\boldsymbol{H}^\mathrm{T}_{\text{ToA}}(i) \boldsymbol{H}_{\text{ToA}}(i)\right)^{-1}}, \label{eqn: GDOP}
\end{equation}   
where $\boldsymbol{H}_{\text{ToA}}(i)$ is obtained from \eqref{eq: wireless distance} and can be written as 
\begin{equation}
     \boldsymbol{H}_{\text{ToA}}(i)=\begin{bmatrix}
         \frac{U_x^{(i)}-P_x^{(1)}+r}{d_{1,i}}, \frac{U_y^{(i)}-P_y^{(1)}+r}{d_{1,i}}\\
         \frac{U_x^{(i)}-P_x^{(2)}+r}{d_{2,i}}, \frac{U_y^{(i)}-P_y^{(2)}+r}{d_{2,i}}\\
         \frac{U_x^{(i)}-P_x^{(3)}+r}{d_{3,i}}, \frac{U_y^{(i)}-P_y^{(3)}+r}{d_{3,i}}
     \end{bmatrix}. \label{eqn: ToA Jocabin}
\end{equation}

It should be noted that this paper emerges as an initial study of new fundamental problems that appear when LTI systems and non-linear systems are connected in cascade. Depending on how the system is implemented and in what way the switches are configurable, $r$ might be known as a priori, in which case its estimation would be redundant. The initial working assumption of the architecture, however, was that there may not be a centrally controllable switch, such that the CU relies on the cumulative signal distortion of the cascaded LTI/non-linear propagation structure to estimate $r$.

\subsection{RU with PAs operating in the linear regime}
Following the discussion in Section~\ref{System overview}, this paper considers PAs working in both linear and non-linear regimes. In this subsection, let us consider PAs working in the linear regime. Under this scenario, the whole RoF system is LTI. Thus, the signal model considers the dispersive channel of PMFs and the wireless channel from the UE to the entry RU. It is assumed that the UE transmits a known sequence of samples, $\boldsymbol{s}=[s_0, \cdots, s_{K-1}]^{\mathrm{T}} \in \mathbb{C}^{K\times1}$ over $K$ frequency points. Initially, the signals undergo the wireless channel between the UE and the RU, which is unknown to the CU.  

Thus, the $k$th element of the input signal at the RoF, $x_k$ with $\boldsymbol{x}=[x_0,x_1,\cdots, x_{K-1}]^{\mathrm{T}}$ (see Fig.~\ref{signal model linear}), can be written as 
\begin{equation}
    x_k=Ae^{-j2\pi f_k \tau}s_k, \label{xk0}
\end{equation}
where $\tau$ can be expressed as 
\begin{equation}
        \tau = \frac{d}{c} + \delta_{\text{t}}, \label{def of tau}
\end{equation}
with $c$ being the speed of light, and $\delta_{\text{t}}$ being the clock offset between the UE and the RoF. Note that distinguishing the effects of $d$ and $\delta_{\text{t}}$ is not possible since they have the same influence over the signals, i.e., phase shift and delay (see \eqref{def of tau}). However, one can rely on the effects of cascaded fibers and PAs to estimate $r$ and $\tau$ separately.

The $k$th element of the output signal at the entry RU, denoted as $y_k^{(0)}$, is given by 
\begin{equation}
    y_k^{(0)}=G Ae^{-j2\pi f_k \tau}s_k + w_k^{(0)},\label{xk1}
\end{equation}
with $w^{(r)}_k \sim \mathcal{CN}(0,\sigma^2)$ being the $k$th AWGN noise component at the $r$th stage. The wireless channel as such does not contribute noise; only the PAs add noise, as they are the only active electronic components. 

In a general form, after propagating through $r$ units (i.e., $r$ PMF segments propagation distance) along the RoF, the received signal at the CU, with $\boldsymbol{y}^{(r)}=[y_0^{(r)}, y_1^{(r)},\cdots,y_{K-1}^{(r)}]^{\mathrm{T}}$, can be written as
\begin{equation}
         y_k^{(r)}=G^{r+1} H_k^{r} Ae^{-j2\pi f_k \tau} s_k + \underbrace{G^{r} H_k^{r} w_k^{(0)}+\cdots + w_k^{(r)}}_{(r+1) \,\text{independent noises}}.\label{CU signal}
\end{equation}
Note that the CU also amplifies the signals. Hence, there are $(r+1)$ PAs and $r$ PMF segments.

The cumulative filtered noise by linear PAs and fibers still follows the Gaussian distribution since the linear transformation does not change the probability distribution. Let $\Tilde{w}_k$ denote the sum of independent Gaussian noise terms, then the variance of $\Tilde{w}_k$ is given by
\begin{equation}
    \mathrm{Var}{(\Tilde{w}_k)}=\sum_{r'=0}^{r} b_k^{r'} \sigma^2. \label{variance of wk}
\end{equation} 
where $b_k \triangleq (G\lvert H_k \rvert)^2.$ Until now, the parameter $r$ takes discrete values, indicating the number of stages that the sub-THz signals propagate in an RoF stripe. Since CRLB only applied to continuous-valued variables, we propose a continuous-value representation of $r$ in Proposition 1. 

\textbf{Proposition 1:} 
\textit{The discrete-valued $r$ in \eqref{variance of wk} can be approximated as a continuous value by splitting one RU into infinitely dense virtual RUs with length $\Delta r\approx0$. The response of one virtual RU is $(G H_k)^{\Delta r}$. By accepting this approximation, the response of one RU is $[(GH_k)^{\Delta r}]^\frac{1}{\Delta r}$, which is equivalent to $ G H_k$. The variance of effective noise at \eqref{variance of wk} can be rewritten as}
\begin{equation}
     \mathrm{Var}{(\Tilde{w}_k)} =\sigma^2 \frac{\ln{b_k}}{b_k-1} \int_0^{r+1} (b_k)^{r'}d r'. \label{continuous variance}
\end{equation}
\textit{Proof: the variance of $\Tilde{w}_k$ in~\eqref{variance of wk} can be re-expressed as the sum of a geometric series as}
\begin{equation}
    \sum_{r'=0}^{r} b_k^{r'} \sigma^2=\frac{1-b_k^{r+1}}{1-b_k} \sigma^2, 
\end{equation}
\textit{which is equal to the result of \eqref{continuous variance} as}
\begin{subequations}
\begin{align}
        \sigma^2\frac{\ln{b_k}}{b_k-1} \int_0^{r+1} (b_k)^{r'}d r'&=\frac{\ln{b_k}}{b_k-1} \frac{b^{r+1}-1}{\ln{b_k}}\sigma^2, \notag \\
        &=\frac{b_k^{r+1}-1}{b_k-1} \sigma^2. \tag{15} \label{proposition}
\end{align}
\end{subequations}
Note that~\eqref{proposition} only holds for $b_k\neq 1, \forall k$. In other words, there are two cases worth being studied. We name them as \textit{flat fiber} and \textit{frequency-selective fiber} cases. This approach was also adopted in~\cite{Henk_GLOBCOM_2023} to calculate the CRLB of a discrete variable. 
\subsubsection{Flat fiber}
Let us consider a special case in which the frequency response of the fiber is totally flat, that is, the case for a system with very limited bandwidth. With the use of PMF, the frequency response can be very flat at 1\,GHz bandwidth. It is assumed that the amplification $G$ is set to compensate for the attenuation, corresponding to $b_k=1, \forall k$. Consequently, the variance of effective noise is 
\begin{equation}
      \boldsymbol{\Sigma} = (r+1) \sigma^2 \boldsymbol{I}_{K},  \label{Sigma flat}
\end{equation}
and the phase shift induced by the fiber grows linearly with the propagation distance $r$. In other words, a flat fiber only supports one-mode transmission within the fiber, corresponding to a single-tap impulse response. It should be noted that the system can not separate the effects of the wireless channel and the RoF based on the phase, but only relies on the noise variance shown in \eqref{Sigma flat} to estimate $r$.

The unknown parameters to be estimated in this problem are $\boldsymbol{\theta}=[A,r,\tau]$, of which the parameters $r$ and $\tau$ lead to a non-linear problem while $A$ is involved linearly. We employ a maximum likelihood (ML) estimator to tackle this estimation problem. First, we define the vector $\boldsymbol{g}\in \mathbb{C}^{K \times 1}$ as 
\begin{equation}
    \boldsymbol{g}=G^{r+1}[e^{-j2\pi f_0 \tau} H_0^{r} s_0,\cdots, e^{-j2\pi f_{K-1} \tau} H_{K-1}^{r} s_{K-1}]^\mathrm{T}. 
\end{equation}
The log-likelihood function can be written as 
\begin{equation}\label{Maximum Likelihood}
\ln\! P(\boldsymbol{y}^{(r)}\lvert \boldsymbol{\theta})\!= \!\ln\!{\big(\pi\mathrm{det}(\boldsymbol{\Sigma})\big)}+(\boldsymbol{y}^{(r)} \! -A\boldsymbol{g})^\mathrm{H} \boldsymbol{\Sigma}^{-1} (\boldsymbol{y}^{(r)}\!-A\boldsymbol{g}).
    \end{equation} 
Since $A$ is linearly involved in the likelihood function and irrelevant to $\boldsymbol{\Sigma}$, the estimation of $A$ can be obtained as
\begin{equation}
\hat{A}=\mathop{\min}_{A}\lvert\lvert \boldsymbol{y}^{(r)}- A \boldsymbol{g} \rvert\rvert^2. 
\end{equation}
It is not difficult to obtain its estimate as
\begin{equation}
\hat{A}= \frac{\boldsymbol{g}^{\mathrm{H}} \boldsymbol{y}^{(r)}}{\lvert\lvert\boldsymbol{g}\rvert\rvert^2}. \label{A_hat}
\end{equation}
After dropping the dependency on $A$ by utilizing $\hat{A}$ and combining with the expression of $\boldsymbol{\Sigma}$ in \eqref{Sigma flat}, we can rewrite the objective function as 
\begin{equation}\label{ML estimator}
    \begin{bmatrix}
   \hat{r}\\
     \hat{\tau}
      \end{bmatrix}\!=\! \mathop{\min}_{r, \tau} \ln {\left((r\!+\!1)K \pi \sigma^2\right)}\!+\frac{\lvert  \lvert \boldsymbol{y}^{(r)}\!-\! \boldsymbol{g} (\boldsymbol{g}^{\mathrm{H}}\boldsymbol{g})^{-1}\boldsymbol{g}^{\mathrm{H}}\boldsymbol{y}^{(r)} \rvert \rvert^2}{(r+1)\sigma^2}.
\end{equation}
Therefore, after a two-dimensional grid search, the estimate $\hat{r}$ as well as $\hat{\tau}$, can be obtained. The estimate of $A$ is calculated as \eqref{A_hat} by plugging in the estimates $\hat{r}$ and $\hat{\tau}.$ 
\subsubsection{Frequency-selective fiber}
A frequency-selective fiber means the magnitude of its frequency response fluctuates over the bandwidth of interest, which makes $b_k$ not equal to one. Without the loss of generality, we assume $b_k \neq 1, \forall k$, which ensures the validity of \eqref{proposition} for all frequencies. This is the case for a system with a large bandwidth, for example, 10\,GHz. Consequently, the variance of effective noise is 
\begin{equation}
    \boldsymbol{\Sigma}=\mathrm{diag}\left\{\frac{b_0^{r+1}-1}{b_0-1} \sigma^2, \frac{b_1^{r+1}-1}{b_1-1} \sigma^2, \cdots,\frac{b_{K-1}^{r+1}-1}{b_{K-1}-1} \sigma^2\right\} \label{Sigma selective}, 
\end{equation}
and the phase shifts introduced by the fiber might grow non-linearly with the propagation distance $r$. Similarly, one can view a \textit{frequency-selective fiber} to support multi-mode transmission, whose corresponding impulse response has multiple taps. 

With the noise covariance being $\eqref{Sigma selective}$, prewhitening is applied by filtering the received signal $\boldsymbol{y}^{(r)}$ and the vector $\boldsymbol{g}$, which can be written as
\begin{subequations}
    \begin{align}
        \Tilde{\boldsymbol{y}}^{(r)}&=\boldsymbol{\Sigma}^{-\frac{1}{2}}\boldsymbol{y}^{(r)}, \notag \\
        \Tilde{\boldsymbol{g}}&=\boldsymbol{\Sigma}^{-\frac{1}{2}}\boldsymbol{g}. \tag{23}
    \end{align}
\end{subequations}
Then, the estimate of $A$ is obtained by \eqref{A_hat}, and the estimates of $r$ and $\tau$ are calculated as \eqref{ML estimator}.  

\subsection{RU with PAs operating in the non-linear regime}
For EE purposes, the PAs are allowed to work in the non-linear regime, as modeled in Section~\ref{System overview}. As previously stated, the UE transmits a known sequence of samples $\boldsymbol{s}=[s_0,\cdots,s_{K-1}]^\mathrm{T}$ over $K$ frequency points to the RoF. We model the input signal in the time domain as
\begin{equation}
    x_n=\frac{1}{K}\sum_{k=0}^{K-1} e^{j2\pi\frac{n}{N} k} A e^{-j 2\pi f_k \tau}s_k,\label{xn0}
\end{equation}
with $x_n \in [x_0,x_1,\cdots, x_{K-1}]^\mathrm{T}.$

The output of the entry RU can be written as  
\begin{equation}
y_{n}^{(0)}=G\Big(x_{n}+\!\!\underbrace{\lambda x_{n} \left\lvert x_{n} \right \rvert^2}_{\text{PA's non-linearity}}\!\!\!\Big)+w_{n}^{(0)}. \label{xn1}
\end{equation}

To obtain the output at the following units, we define the function $f(\cdot)$, which combines the effects of a non-linear PA and one segment of PMF, and can be expressed as
\begin{equation}
        f(y_{n}^{(0)}) = G\left(\sum_{l=0}^{L-1} \beta_l y_{n-l}^{(0)}+\lambda \sum_{l=1}^{L-1} \beta_l y_{n-l}^{(0)} \left\lvert \sum_{l=1}^{L-1} \beta_l y_{n-l}^{(0)} \right\rvert^2\right). \label{ffunction} 
\end{equation}
If $f(\cdot)$ in \eqref{ffunction} takes a vector as input, it operates element-wise. 

The signal $\boldsymbol{y}^{(0)}$ would undergo $r$ PAs and $r$ segments of PMF to reach the CU. In other words, the signal $\boldsymbol{y}^{(0)}$ would undergo a recursive process $r$ times. Let $f^{r}(\boldsymbol{y}^{(0)})$ be the overall function, which can be expressed as 
\begin{equation}
f^{r}(\boldsymbol{y}^{(0)})=\underbrace{f(f(f(...(\boldsymbol{y}^{(0)}))))}_{r \, \text{times}}.
\end{equation} 
Therefore, the received signal at the CU can be expressed as $\boldsymbol{y}^{(r)}= f^{r}\left(\boldsymbol{y}^{(0)}\right)+\boldsymbol{w}^{(r)}$, where all the unknown parameters are interwoven with each other. As a result, we cannot apply the same estimator proposed in the linear regime (\eqref{A_hat} and \eqref{ML estimator}). Alternatively, we can employ an NLS framework. 

With PAs operating in the non-linear regime, an NLS framework may not serve as an optimal estimator as the noise distribution is transformed from a Gaussian distribution into another probability distribution. However, the output of an NLS framework will not deviate significantly from the optimum. Following the NLS framework, the objective function w.r.t. the unknown parameter $A,r,$ and $\tau$ is given by 
    \begin{equation}
        \begin{bmatrix}
        \hat{A}\\
        \hat{\tau}\\
        \hat{r}
    \end{bmatrix}=\mathop{\min}_{A,\tau,r} \bigg\lvert\bigg\lvert\boldsymbol{y}^{(r)}-f^{r}\left(G\left(x_n+\lambda x_n \left\lvert x_n \right \rvert^2\right)\right)\bigg\rvert\bigg\rvert^2,\label{ml}
\end{equation}
where a three-dimensional grid search is needed to obtain the estimates. To solve this problem, we employ a particle swarm optimization (PSO) algorithm described in Algorithm~\ref{algo:PSO}, to reduce the complexity compared to a three-dimensional grid search. 

Compared to the ML estimator in \eqref{Maximum Likelihood}, the employed NLS framework in the non-linear regime disregards the information contained in the noise covariance. 

\subsection{Positioning of the UE}
According to \eqref{eq: wireless distance}, the distance between the $i$th UE and the entry RUs is written as $\boldsymbol{d}\triangleq[d_{1,i},d_{2,i},d_{3,i}]^\mathrm{T}$. 
The ToA from the UE to the entry RUs is interwoven with a clock offset as defined in \eqref{def of tau}, which can be viewed as a bias. The position of the UE $U_x^{(i)}, U_y^{(i)}$ is estimated through the estimated $\boldsymbol{\tau}=(1/c)\boldsymbol{d}+\mathbf{1}_3\delta_t$ at three RoF stripes and can be written as 
\begin{equation}
    \begin{bmatrix}
   \hat{U_x}^{(i)}\\
   \hat{U_y}^{(i)}
      \end{bmatrix}=\mathop{\min}_{U_x^{(i)},U_y^{(i)},\delta_{\text{t}}} \lvert\lvert \boldsymbol{\tau}-\frac{\boldsymbol{d}}{c}-\mathbf{1}_3\delta_{\text{t}}\rvert \rvert^2, \label{estimate of position} 
\end{equation}
where the estimation of $\delta_{\text{t}}$ is a linear model with the rest of quantities. Let $\boldsymbol{R}\triangleq \boldsymbol{\tau}-\frac{\boldsymbol{d}}{c}$. It is not difficult to write the estimate of $\delta_{\text{t}}$ as 
\begin{equation}
    \hat{\delta_{\text{t}}}=\frac{\mathbf{1}_3^\mathrm{T}\boldsymbol{R}}{\mathbf{1}_3^\mathrm{T}\mathbf{1}_3}.
\end{equation}
Plugging in $\hat{\delta_{\text{t}}}$ back to \eqref{estimate of position}, the estimate of the UE positions can be written as 
\begin{equation}
    \begin{bmatrix}
   \hat{U_x}^{(i)}\\
   \hat{U_y}^{(i)}
      \end{bmatrix}=\mathop{\min}_{U_x^{(i)}, U_y^{(i)}} \lvert\lvert \boldsymbol{R}-\mathbf{1}_3 (\mathbf{1}_3^{\mathrm{T}}\mathbf{1}_3)^{-1}\mathbf{1}_3^{\mathrm{T}}\boldsymbol{R}\rvert \rvert^2. 
\end{equation}
\subsection{The effect of the non-ideal PA}\label{sec: non-ideal PA}
The analysis of this section is based on ideal PA assumptions, i.e., all PAs have identical and constant characteristics. Although this assumption is ideal, a small deviation from it does not lead to significant effects on the estimators. Practically, the noise variance and PA gain show little variation with the room temperature (much less than 2\,dB gain variation in \cite{zhou2022chip}). The cumulative impact of cascading PAs is much more dominant than that of varying individual PA characteristics. For example, two PAs in a cascade doubles the noise variance and squares the gain, compared to a single PA.

\section{Cramér-Rao Lower Bound}\label{Cramer-Rao Lower bound}
CRLB is a very useful tool for evaluating the performance of any unbiased estimator, as it provides a measure of the lower limit of the variance. However, the CRLB for the case of PAs operating in the non-linear regime leads to a cumbersome problem due to the presence of iterative non-linearity in the signal model. Alternatively, this section introduces the CRLB of the proposed ML estimator with PAs operating in the linear regime as a benchmark of estimation accuracy.

For simplicity, we express the complex valued $A$ and $H$ with their magnitudes $\lvert A \rvert$  and $\lvert H_k \rvert$ and phases $\phi$ and $\psi_k$, i.e., $A=\lvert A \rvert e^{j\phi}$ and $H=\lvert H_k \rvert e^{j\psi_k}$. Hence, the unknown parameters are defined as $\boldsymbol{\theta} \triangleq [\lvert A \rvert, \phi, \tau, r]^{\mathrm{T}}. $ The definition for the CRLB is given by 
\begin{equation}
    \left[\mathrm{Var}(\boldsymbol{\hat{\theta}_{i}})\right] \geq \left[\boldsymbol{I}^{-1}(\boldsymbol{\theta})\right]_{i,i}, \label{CRLB}
\end{equation}
where $\boldsymbol{I}(\boldsymbol{\theta})$ is the Fisher information matrix (FIM) for the vector $\boldsymbol{\theta}$ and is defined as
\begin{subequations}
\begin{align}
 \left[\boldsymbol{I}(\boldsymbol{\theta})\right]_{i,j} &= \mathrm{tr}\left\{\boldsymbol{C}(\boldsymbol{\theta})^{-1}\frac{\partial \boldsymbol{C}(\boldsymbol{\theta})}{\partial\boldsymbol{\theta}_{i}} \boldsymbol{C}(\boldsymbol{\theta})^{-1} \frac{\partial \boldsymbol{C}(\boldsymbol{\theta})}{\partial\boldsymbol{\theta}_{j}}\right\} \notag \\
 &+2\Re\left\{\frac{\partial \boldsymbol{\mu(\theta)}^\mathrm{H}}{\partial \theta_i }\boldsymbol{C}(\boldsymbol{\theta})^{-1} \frac{\partial \boldsymbol{\mu(\theta)}}{\partial \theta_j} \right\}, \label{FIM definition} \tag{33}
\end{align}
\end{subequations}
which reveals the amount of information associated with each pair of unknown parameters. The FIM in \eqref{FIM definition} is valid only for Gaussian models. One important characteristic of FIM is that every FIM is positive semi-definite, indicating the information content is non-negative and ensuring its eigenvalues are non-negative. Moreover, a positive definite FIM indicates the parameters are identifiable, and a singular FIM indicates the parameters are non-identifiable. 

As mentioned in Section \ref{PF and SM}, there are two expressions for the variance of effective noise as shown in \eqref{Sigma flat} and in \eqref{Sigma selective}. Naturally, the CRLB needs to be calculated for both flat fiber and frequency-selective fiber cases. Detailed calculations can be found in the Appendix~\ref{append}.
\subsection{CRLB for flat fiber}
Now, let us start with the flat fiber case, whose $\boldsymbol{C}(\boldsymbol{\theta})$ and $\boldsymbol{\mu}(\boldsymbol{\theta})$ are defined as
\begin{subequations}
    \begin{align}
       \boldsymbol{C}(\boldsymbol{\theta}) &= (r+1) \sigma^2 \boldsymbol{I}_{K} \notag \\ 
    \boldsymbol{\mu(\theta)} &= \left[
      G^{r+1}\lvert A \rvert \lvert H_0 \rvert^r s_0 e^{j\left(\phi + r \psi_{0}-2\pi f_0 \tau \right)}, \ \cdots, \right. \notag\\
      & \qquad \left. G^{r+1} \lvert A \rvert \lvert H_{K-1}\rvert^r s_{K-1} e^{j\left(\phi+ r \psi_{K-1}-2\pi f_{K-1} \tau\right)} \right]^\mathrm{T}. \tag{34} \label{mean vector}
    \end{align}
\end{subequations}

According to the symmetry property, $\left[\boldsymbol{I}(\theta)\right]_{i,j}=\left[\boldsymbol{I}(\theta)\right]_{j,i}$ is always valid so that there are 10 unique entries in each FIM.  
Regarding the \textit{flat fiber}, the elements of the FIM are 
\begin{subequations}
    \begin{align}
    \left[\boldsymbol{I}(\theta)\right]_{\lvert A \rvert,\lvert A \rvert}=& 2\frac{G^{2} }{(r+1)\sigma^2} \sum_{k=0}^{K-1}\lvert s_k \rvert^2, \label{FIM amplitude flat} \tag{35a} \\
    \left[\boldsymbol{I}(\theta)\right]_{\lvert A \rvert,\phi}= &0, \label{FIM flat A phi} \tag{35b} \\
     \left[\boldsymbol{I}(\theta)\right]_{\lvert A \rvert,\tau}=& 0,\label{FIM flat A tau} \tag{35c} 
     \end{align}
\end{subequations}   
 \begin{subequations}
     \begin{align}     
    \left[\boldsymbol{I}(\theta)\right]_{\lvert A \rvert,r}=& 0, \label{FIM flat A r} \tag{35d}\\
    \left[\boldsymbol{I}(\theta)\right]_{\phi,\phi}=& 2\frac{G^2 \lvert A \rvert^2}{(r+1)\sigma^2} \sum_{k=0}^{K-1}  \lvert s_k \rvert^2, \label{FIM flat phi phi}\tag{35e}\\
    \left[\boldsymbol{I}(\theta)\right]_{\phi,\tau}=& -\frac{ 4\pi G^2 \lvert A \rvert^2}{(r+1)\sigma^2} \sum_{k=0}^{K-1} f_k   \lvert s_k \rvert^2,\label{FIM flat phi tau} \tag{35f} \\
    \left[\boldsymbol{I}(\theta)\right]_{\phi,r}=& 2\frac{G^{2} \lvert A \rvert^2 }{(r+1)\sigma^2} \sum_{k=0}^{K-1} \lvert s_k \rvert^2 \psi_k ,\label{FIM flat phi r} \tag{35g}\\
    \left[\boldsymbol{I}(\theta)\right]_{\tau,\tau}=&\frac{8\pi^2 G^2 \lvert A \rvert^2 }{(r+1)\sigma^2} \sum_{k=0}^{K-1} f_k^2 \lvert s_k\rvert^2,\label{FIM flat tau tau} \tag{35h}\\ 
    \left[\boldsymbol{I}(\theta)\right]_{r,r}=&\frac{K}{(r+1)^2}+2\frac{G^{2} \lvert A\rvert^2}{(r+1) \sigma^2} \sum_{k=0}^{K-1} \lvert s_k\rvert^2 \psi_k^2, \label{FIM flat r r} \tag{35i}\\
    \left[\boldsymbol{I}(\theta)\right]_{\tau,r}=& -\frac{4\pi G^2 \lvert A \rvert^2}{(r+1)\sigma^2} \sum_{k=0}^{K-1}\lvert s_k \rvert^2 f_k \psi_k. \tag{35j}\label{FIM flat r tau}  
\end{align}
\end{subequations}
\begin{figure*}[ht]
    \centering
    \begin{subfigure}[t]{0.47\linewidth}
        \centering
        \includegraphics[width=1\linewidth]{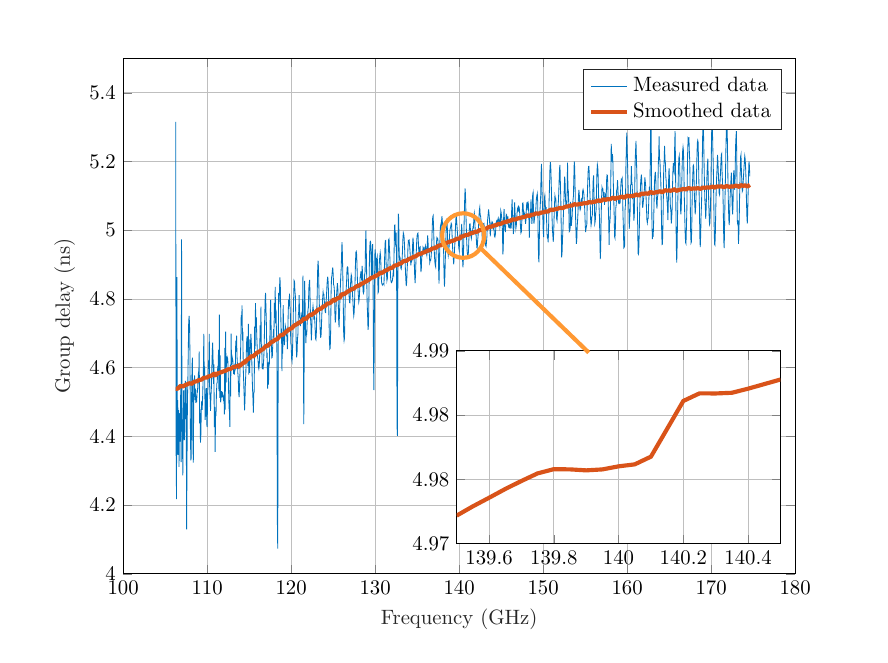}
        \caption{}
        \label{gd}
    \end{subfigure}
    \hfill
    \begin{subfigure}[t]{0.47\linewidth}
        \centering
        \includegraphics[width=1\linewidth]{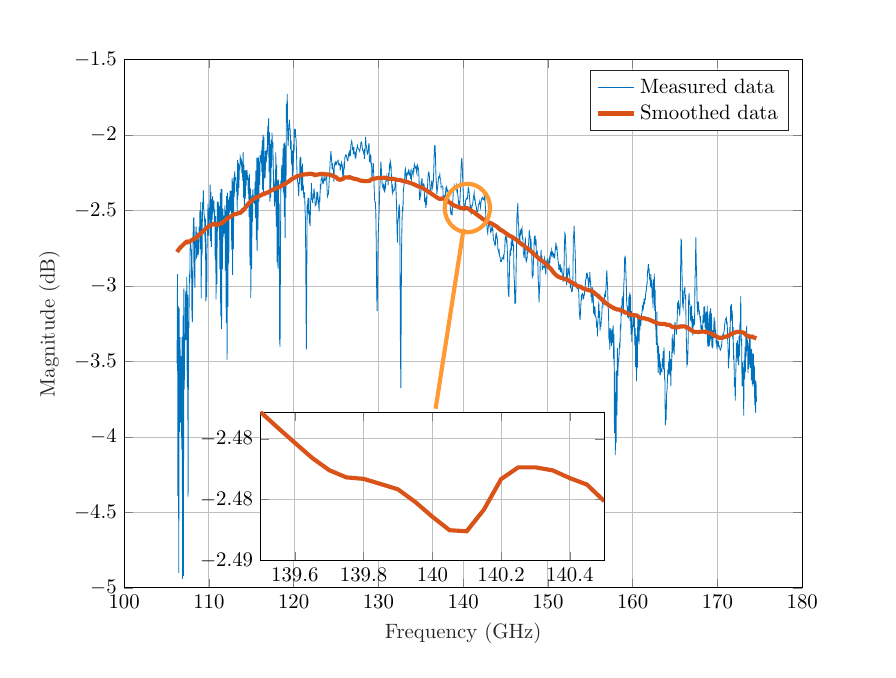}
        \caption{}
        \label{pl}
    \end{subfigure}
    
    \caption{Measured characteristics of a 1-meter PMF made of high-density polyethylene with a solid rectangular cross-section in the D-band (110\,GHz to 170\,GHz) \cite{Frida}: (a) Group delay; (b) Magnitude.}
    \label{fig:combined}
\end{figure*}

\textbf{Discussions:} The main takeaways from the CRLB calculation, from \eqref{FIM amplitude flat} to \eqref{FIM flat r tau}, are: \begin{enumerate}[label=\roman*.] 
    \item The estimate of $\lvert A \rvert$ is solely dependent on the received signal-to-noise ratio (see \eqref{FIM amplitude flat}).  
    \item As shown in \eqref{FIM flat phi tau}, \eqref{FIM flat phi r}, and \eqref{FIM flat r tau}, there is a coupling between the estimates of $\tau$, $r$, and $\phi$, which is proved to make the FIM close to a singular matrix through numerical results in the following section. 
    \item The phase shifts introduced by the fiber are interwoven with the phase shifts introduced from the wireless channel and the phase offset, making it impossible for the system to identify the propagation distance $r$ based on the phase information.
    \item According to \eqref{FIM flat r r}, the system can rely on the overall noise variance to estimate the propagation distance, as it cumulates linearly with $r$. 
\end{enumerate}

\subsection{CRLB for frequency-selective fiber}
For the frequency-selective fiber case, the noise variance is defined as 
\begin{subequations}
\begin{align}
     \boldsymbol{C}(\boldsymbol{\theta}) &= \mathrm{diag}\left(\sigma^2 \frac{\ln{b_0}}{b_0-1} \int_0^{r+1} (b_0)^{r'}d r', \cdots, \right. \notag\\
       & \qquad \left. \sigma^2 \frac{\ln{b_{K-1}}}{b_{K-1}-1} \int_0^{r+1} (b_{K-1})^{r'}d r' \right), \notag \\
     \left[\boldsymbol{C^{-1}}(\boldsymbol{\theta})\right]_{k,k}&=\frac{b_k-1}{\sigma^2(b_k^{r+1}-1)}, \notag \\
        \left[\frac{\partial \boldsymbol{C(\theta)}}{\partial r} \right]_{k,k}&= \frac{\sigma^2b_k^{r+1} \ln{b_k}}{b_k-1}, \label{mean vector selective} \tag{36} 
\end{align}
\end{subequations}
and $\boldsymbol{\mu(\theta)}$ is the same as \eqref{mean vector}. Similarly, the entries in the FIM can be obtained as
\begin{subequations}
    \begin{align}
    \left[\boldsymbol{I}(\theta)\right]_{\lvert A \rvert,\lvert A \rvert}=& 2G^2\sum_{k=0}^{K-1}\frac{(b_k-1)b_k^r\lvert s_k \rvert^2}{\sigma^2 (b_k^{r+1}-1)},\label{FIM amplitude selective} \tag{37a}\\
    \left[\boldsymbol{I}(\theta)\right]_{\lvert A \rvert,\phi}=& 0,\tag{37b} \\
    \left[\boldsymbol{I}(\theta)\right]_{\lvert A \rvert,\tau}=& 0,\tag{37c} 
     \end{align}
 \end{subequations}    
 \begin{subequations}
 \begin{align} 
    \left[\boldsymbol{I}(\theta)\right]_{\lvert A \rvert,r}\!=&2 G^{2} \lvert A \rvert \!\sum_{k=0}^{K-1}\!\frac{(b_k-1) b_k^{r} \lvert s_k\rvert^2 \ln{\sqrt{b_k}}}{\sigma^2(b_k^{r+1}-1)}, \tag{37d} \label{FIM amplitude r selective} \\
    \left[\boldsymbol{I}(\theta)\right]_{\phi,\phi}=& 2 G^{2} \lvert A \rvert^2\sum_{k=0}^{K-1} \frac{(b_k-1)b_k^{r}\lvert s_k \rvert^2}{\sigma^2(b_k^{r+1}-1)},\tag{37e} \label{FIM selective phi phi}\\
    \left[\boldsymbol{I}(\theta)\right]_{\phi,\tau}\!=&\! -4 \pi G^{2} \lvert A \rvert^2\!\sum_{k=0}^{K-1}\!\frac{(b_k-1) f_k b_k^{r} \lvert s_k \rvert^2}{\sigma^2(b_k^{r+1}-1)},\tag{37f} \label{FIM selective phi tau} \\
    \left[\boldsymbol{I}(\theta)\right]_{\phi,r}=& 2 G^{2} \lvert A \rvert^2 \sum_{k=0}^{K-1} \frac{(b_k-1) b_k^{r} \lvert s_k \rvert^2 \psi_k}{\sigma^2(b_k^{r+1}-1)}, \tag{37g} \label{FIM selective phi r}\\
    \left[\boldsymbol{I}(\theta)\right]_{\tau,\tau}=&8\pi^2 G^{2} \lvert A \rvert^2 \sum_{k=0}^{K-1}\frac{ (b_k-1) f_k^2 b_k^{r} \lvert s_k\rvert^2}{\sigma^2(b_k^{r+1}-1)} \tag{37h} \label{FIM selective tau tau}, \\
    \left[\boldsymbol{I}(\theta)\right]_{\tau,r}\!=\!&- 4\pi G^{2}\lvert A \rvert^2\!\sum_{k=0}^{K-1}\frac{(b_k-1) b_k^{r} \lvert s_k \rvert^2 f_k \psi_k}{\sigma^2(b_k^{r+1}-1)} \tag{37i} \label{FIM selective r tau},\\
    \left[\boldsymbol{I}(\theta)\right]_{r,r}=&2 G^{2} \lvert A\rvert^2 \!\!\!\sum_{k=0}^{K-1}\!\! \frac{\lvert s_k\rvert^2 (b_k\!-\!1) b_k^{r}\! \left( (\ln{\!\sqrt{b_k}})^2\!\!+\!\psi_k^2 \right)}{\sigma^2(b_k^{r+1}-1)} \notag \\
       & +\sum_{k=0}^{K-1}\left(\frac{b_k^r\ln{b_k}}{1-b_k^{r+1}}\right)^2. \tag{37j} \label{FIM selective r r}
\end{align}
\end{subequations}
\textbf{Discussion:} Based on the CRLB obtained for the frequency-selective fiber, one can observe \begin{enumerate}[label=\roman*.] 
    \item With the use of frequency-selective fiber, a coupling between $\lvert A \rvert$ and $r$ appears as illustrated in \eqref{FIM amplitude r selective}, which comes from the varied magnitude of the frequency response of the PMF.
    \item As illustrated in \eqref{FIM selective r r}, the system can rely on the shape of the spectrum of the received signal at CU, while it can only rely on the noise variance with a flat fiber (see \eqref{FIM flat r r}).
    \item The coupling between $\phi$, $\tau$, and $r$ still exists (see \eqref{FIM selective phi r}, \eqref{FIM selective phi tau}, and \eqref{FIM selective r tau}). 
\end{enumerate}

\section{Simulation results}\label{simulation results}
In this section, we first validate the proposed NLS estimator, relying on the measured channel of a segment of PMF. Then, we validate the proposed CRLB with two channels generated under a constraint on the total energy. Moreover, we apply the proposed positioning algorithm in an indoor scenario.
\subsection{Validation of the proposed estimators with measured fiber channel characteristics}
In this section, we illustrate the performance obtained by the proposed estimators with the fiber channel measurements presented in~\cite{Frida}. The channel measurement data in \cite{Frida} was obtained by transmitting sub-THz (D-band) signals over a one-meter long high-density polyethylene PMF with a solid rectangular cross-section of $1$\,mm$\times 2$\,mm and a density
of $0.93$\,g/cm$^3$. These measurements inherently include several imperfections, such as noise, interference, and measurement errors. Therefore, a median filtering with a window size of 300 is implemented through MATLAB to smooth the data in group delay and magnitude, separately. Both the smoothed data and the original data are shown in Figs.~\ref{gd} and~\ref{pl}.

The measured data consists of channel characteristics at $K$ discrete frequencies, encompassing both magnitudes ($\lvert \boldsymbol{H}\rvert=\big[\lvert H_0\rvert, \cdots, \lvert H_{K-1} \rvert\big]^\mathrm{T}$) and group delay. Subsequently, the phase response $\boldsymbol{\psi}=[\psi_0, \cdots, \psi_{K-1}]^\mathrm{T}$ was obtained by integrating the group delay on the discrete frequencies. Then, the transfer function can be expressed as
\begin{equation}
 H_k= \lvert H_k\rvert \exp{(j \psi_k)}. \tag{38}\label{MeasuredH}
\end{equation} The employed LoS model comprises free-space path loss and a large-scale fading component, which can be written as \begin{equation}
        \lvert A \rvert=G_t G_r (\frac{\lambda}{2\pi d})^2+\zeta, \label{Magnitude of A} \tag{39} \end{equation}
with $\lambda$ being the wavelength, $d$ being defined in \eqref{eq: wireless distance}, $G_t$ and $G_r$ being the transmitter gain, and receiver gain respectively. The large-scale fading component $\zeta$ follows a log-normal distribution with a standard deviation of 2\,dB. This model can be replaced by other developed LoS models for sub-THz propagation, such as the 3GPP model in \cite{3gpp}. In our simulation, $G_t$ is 36 dB, and $G_r$ is also 36 dB. Assuming the use of state-of-the-art patch antennas with 9 dBi gain~\cite{Yuyan_Eucap_2024, Yuyan_Eucap_2025}, a 22 $\times$ 22 array is required, which translates to 4 cm $\times$ 4 cm antenna size. A horn antenna with equivalent gain can also be used. 

\begin{figure}[t]
    \centering
    \includegraphics[width=1\linewidth]{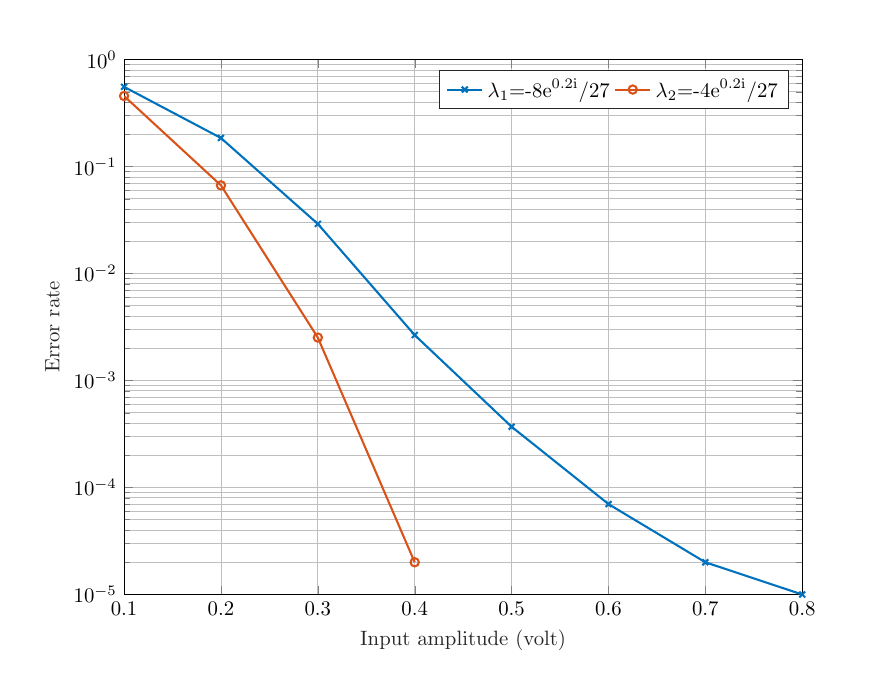}
    \caption{The error rate of $\hat{r}\neq r$.}
    \label{r_non-linear}
\end{figure}
Monte Carlo simulations were performed with an RoF system comprising five RUs and BUs, and one CU. Between every two units, a fiber with the frequency response as expressed in \eqref{MeasuredH} serves as the connection. Distortion incurred by PAs was illustrated in Fig.~\ref{PA charact}. The UE transmits a known sequence of samples using a quadrature phase shift keying waveform with a unit amplitude. The UE accesses the RoF system through the third RU. The channel response shown in Fig.~\ref{pl} corresponds to a spectrum with 1\,GHz bandwidth centered around 140\,GHz, from which the magnitude in a one-meter fiber is approximately -2.48\,dB. Therefore, $G$ is set to 2.48\,dB for PAs. The parameters in \eqref{Magnitude of A} were set to ensure the input amplitude is in a certain regime (linear or non-linear) of PAs.

The proposed NLS estimator was implemented in Matlab, and the results are shown in Fig.~\ref{r_non-linear}. It is observed that the error rate would decrease to a level of $10^{-5}$ with a proper input amplitude, which indicates that the proposed algorithm is able to accurately estimate the propagation distance even when PAs work in the non-linear regime. Through comparison between the estimation accuracy with two non-linear factors, one can observe an accuracy loss with a higher non-linear factor. 

\subsection{Validation of CRLB with assumed fiber channels}
\begin{figure}[t]
    \centering
    \includegraphics[width=1\linewidth]{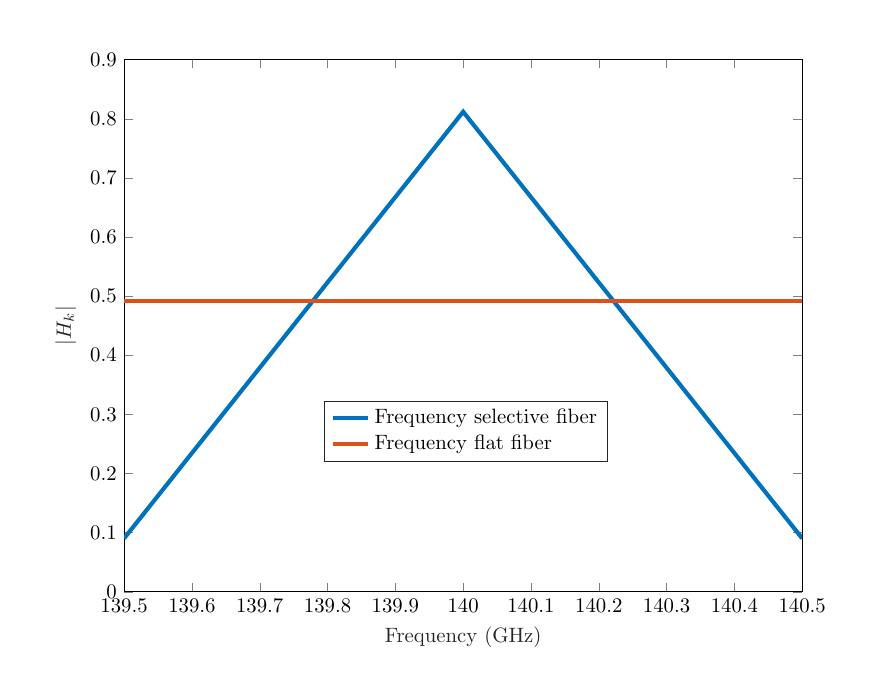}
    \caption{The magnitude of the frequency response for the frequency flat fiber and the frequency-selective fibers.}
    \label{fiber_channels}
\end{figure}
\begin{figure}[t]
        \centering
        \includegraphics[width=0.9\linewidth]{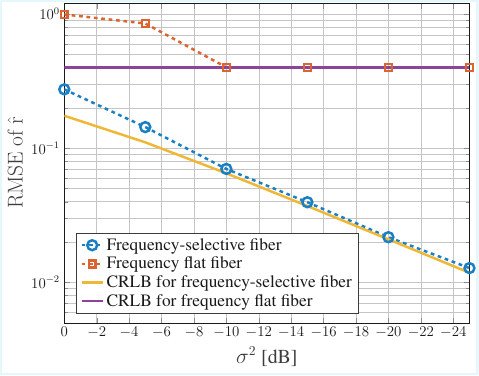}
        \caption{RMSE and CRLB of $\hat{r}$, with frequency flat and frequency-selective fibers.}
        \label{rmse_r}
    \end{figure}
    \begin{figure}[t]
        \centering
        \includegraphics[width=0.9\linewidth]{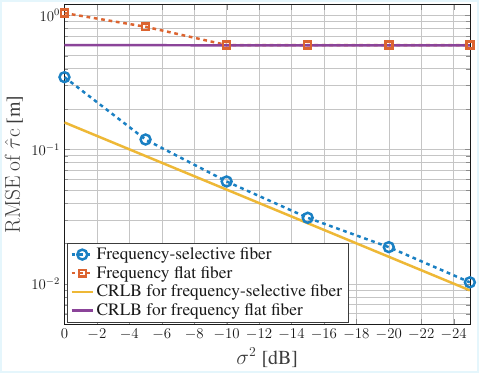}
        \caption{RMSE and CRLB of $\hat{\tau}$, with frequency flat and frequency-selective fibers.}
        \label{rmse_tau}
    \end{figure}
To validate the CRLB calculated, extensive Monte-Carlo simulations were performed and based on two artificial channels shown in Fig.~\ref{fiber_channels}, corresponding to a flat fiber (red curve) and a frequency-selective fiber (blue curve). Each channel contains the same amount of energy, i.e., $\sum_{k=0}^{K-1} \lvert H_k \rvert^2 = \mathcal{E}$. It should be emphasized that the shape of the spectrum of a frequency-selective fiber does not impose any deviation in the conclusions of this paper. The corresponding root mean squared error (RMSE) of target parameters was calculated and plotted in Figs.~\ref{rmse_r}, and ~\ref{rmse_tau}.  

The comparison between the CRLB w.r.t. $\hat{r}$ with a flat fiber and that with a non-flat fiber indicates that the estimation accuracy is enhanced with a frequency-selective fiber, which proves the fact that the system can extract additional information from the spectrum of the received signal. Another observation from Fig.~\ref{rmse_r} is a flat bound with a flat fiber. This understanding can be acquired from \eqref{FIM flat r r}, where the CU estimates the noise variance and determines its quantitative relationship with the noise variance $\sigma^2 $ at each PA. 
Therefore, this bound is independent of the exact value of $\sigma^2$.  The ML estimator achieves an RMSE equal to the CRLB, indicating that it attains optimal performance. Similar conclusions can be drawn from Fig~\ref{rmse_tau}. It is worth noting that the gap between the RMSE of certain estimates and the corresponding CRLB does not necessarily indicate suboptimality, as the simulation grids are inherently constrained.

\begin{figure}[t]
        \centering
        \includegraphics[width=0.9\linewidth]{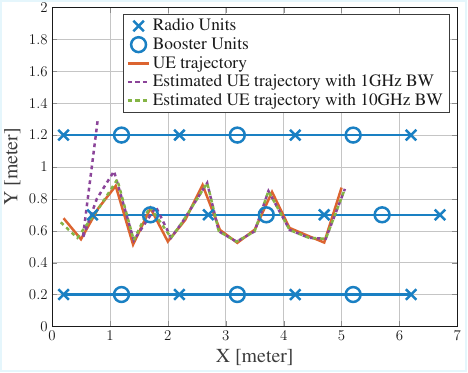}
     \caption{Floorplan of an indoor scenario with three RoF stripes, a UE trajectory, and the estimated UE trajectory with 1\,GHz and 10\,GHz bandwidth.}
        \label{positioning_twc}
    \end{figure}

\begin{figure*}[t]
    \centering
    \begin{subfigure}[b]{0.48\textwidth}
        \centering
        \includegraphics[width=1\linewidth]{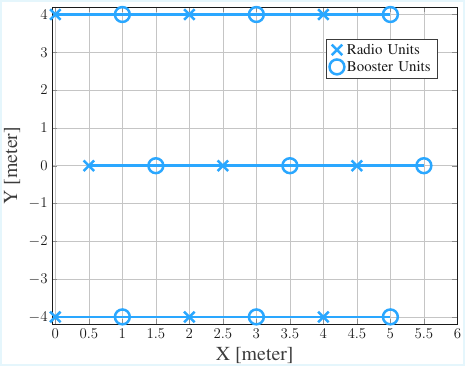}
        \caption{\textcolor{blue}{}}
        \label{fig:parallel topology}
    \end{subfigure}
    \hfill
    \begin{subfigure}[b]{0.48\textwidth}
        \centering
       \includegraphics[width=1\linewidth]{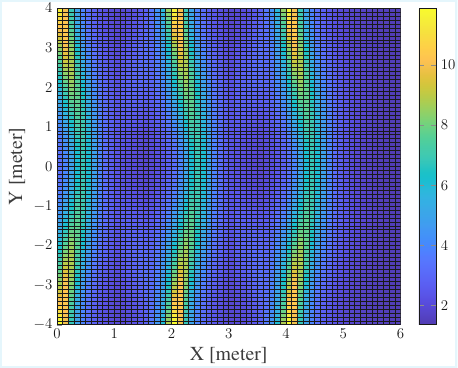}
        \caption{\textcolor{blue}{}}
        \label{fig:parallel gdop}
    \end{subfigure}

    \vspace{0.5cm}

    \begin{subfigure}[b]{0.48\textwidth}
        \centering
        \includegraphics[width=1\linewidth]{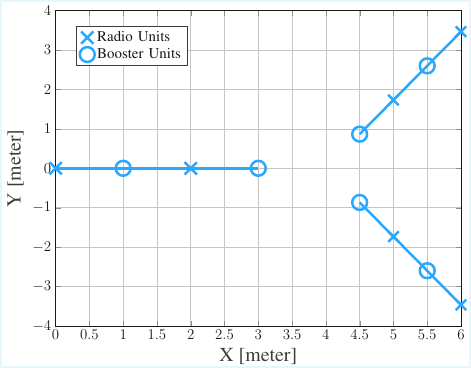}
        \caption{\textcolor{blue}{}}
        \label{fig:star topology}
    \end{subfigure}
    \hfill
    \begin{subfigure}[b]{0.48\textwidth}
        \centering
        \includegraphics[width=1\linewidth]{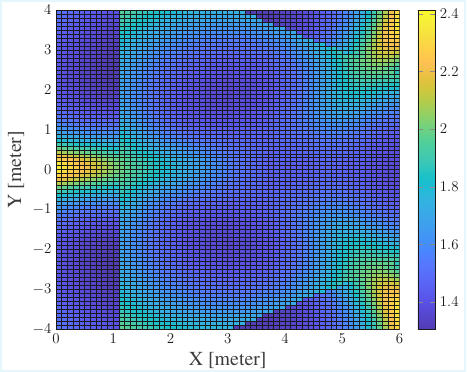}
        \caption{\textcolor{blue}{}}
        \label{fig:star gdop}
    \end{subfigure}

    \caption{Comparison of RU deployments in parallel and star topologies: (a) parallel-topology network, (b) GDOP distribution for the parallel topology, (c) star-topology network, and (d) GDOP distribution for the star topology.}
    \label{GDOP}
\end{figure*}

\subsection{Use case: Indoor positioning}\label{Use case}
A pertinent use case for the proposed estimation framework on the considered RoF system is indoor positioning. This low-cost design allows the operator to densely deploy RoF stripes for extreme data rate and good coverage. Here, we assume a UE moves along a trajectory in an indoor scenario illustrated in Fig.~\ref{indoor scenario}, where three RoF stripes are serving the UE simultaneously. To ensure fairness and avoid potential biases, no additional assumptions about the antennas are made. Instead, it is simply assumed that the UE is consistently served by three RUs located at three RoF stripes. The height of the UE is assumed to be consistently 1.5 meters below the ceiling at all positions, with a uniform spacing of 1 meter between adjacent units in stripes and a spacing of 0.5 meters between adjacent RoF stripes.

The simulations were carried out using bandwidths of 1\,GHz and 10\,GHz, corresponding to delay resolutions of 0.3 meters and 0.03 meters, respectively. Fig~\ref{positioning_twc} shows the estimated UE trajectories with different bandwidths. The RMSE of the estimated positions, obtained through Monte Carlo simulations, is 0.04 meters for 10\,GHz and 0.5 meters for 1\,GHz, both closely approaching their respective delay resolutions. The GDOP quantifies the influence of the geometry on the positioning performance. The proposed positioning algorithm is based on ToA with bias, whose GDOP can be calculated according to \eqref{eq: wireless distance}-\eqref{eqn: GDOP}. Stripe deployments can follow different network topologies, such as the parallel topology, which is the focus of this paper, or a star topology (see Figs.~\ref{fig:parallel topology} and \ref{fig:star topology}). For consistency, both topologies are deployed to cover the same service area; however, one should notice that the star topology deploys fewer RUs than the parallel topology. Figs.~\ref{fig:parallel gdop} and \ref{fig:star gdop} illustrate the GDOP distribution for both topologies. In Figs.~\ref{fig:parallel gdop} and ~\ref{fig:star gdop}, the high GDOP in certain positions is mainly attributed to the alignment of two RUs along a line, which results in an unfavorable geometry and reduced positioning performance. Additionally, the star network topology exhibits greater geometric diversity than the parallel network topology. A topic for future research can be to investigate optimal RU deployments by jointly considering the positioning and communication performance. 

\begin{algorithm}[hbt!]
\small
\setstretch{1.15}
\caption{Particle Swarm Optimization Algorithm}
\label{algo:PSO}

\SetKwFunction{Cost}{Cost}
\SetKwProg{Fn}{Function}{:}{End Function}

\Fn{\Cost{$\boldsymbol{\theta}, \lambda, G, \boldsymbol{s}, f(x), \boldsymbol{y}^{(r)}$}}{
     $x_n=\frac{1}{K}\sum_{k=0}^{K-1} e^{j2\pi\frac{n}{N} k} A e^{-j 2\pi f_k \tau}s_k;$\
      
    $c \gets$ $\bigg\lvert\bigg\lvert\boldsymbol{y}^{(r)}-f^{r}\left(G\left(x_n+\lambda x_n \left\lvert x_n \right \rvert^2\right)\right)\bigg\rvert\bigg\rvert^2\;$
    
    \Return $c$\;
}
\SetKwFunction{PSO}{PSO}
\SetKwProg{Fn}{Function}{:}{End Function}
\Fn{\PSO{$\lambda, G, H, \boldsymbol{s}, \boldsymbol{y}^{(r)}, f(x)$}}{
    $m \gets 50$\ \tcp*[r]{Set the \#iterations}
    $p \gets 100$\ \tcp*[r]{Set particle size}
    $w_1 \gets 1.5$\ \tcp*[r]{personal best pos.Weights}
    $w_2 \gets 1.5$\ \tcp*[r]{global best pos. Weights}
    $w_{\text{ine}} \gets$ 0.3\ \tcp*[r]{Set inertia weight}
    $n \gets$ 4\ \tcp*[r]{\#variables}
    $c_{\text{gbest}} \gets 0$\, $v \gets 0$ \tcp*[r]{Initialize global cost}
    $\theta_{\text{min}} \gets [\lvert A \rvert_{\text{min}},\phi_{\text{min}},\tau_{\text{min}},r_{\text{min}}]$\; 
    $\theta_{\text{max}} \gets [\lvert A \rvert_{\text{max}},\phi_{\text{max}},\tau_{\text{max}},r_{\text{max}}]$\; 
    \For{$i \gets 1$ \textbf{to} $p$}{
        \For{$j \gets 1$ \textbf{to} $n$}{
            $\boldsymbol{\theta}(i, j) \gets (\boldsymbol{\theta}_{\text{max}}(j) - \boldsymbol{\theta}_{\text{min}}(j)) * \text{rand(1)} + \boldsymbol{\theta}_{\text{min}}(j)$\tcp*[r]{Set the initial values}
            $\boldsymbol{\theta}_{\text{aux}}(j) \gets \boldsymbol{\theta}(i, j)$\;
        }
        $\boldsymbol{c}(i) \gets \Cost(\boldsymbol{\theta}_{\text{aux}}, \lambda, G, \boldsymbol{s}, f(x), \boldsymbol{y}^{(r)})$\;
        \If{$\boldsymbol{c}(i) < c_{\mathrm{gbest}}$}{
            $c_{\text{gbest}} \gets \boldsymbol{c}(i)$\;
            \For{$j \gets 1$ \textbf{to} $n$}{
                $\boldsymbol{\theta}_{\text{gbest}}(j) \gets \boldsymbol{\theta}(i, j)$\tcp*[r]{Iterate the global best cost}
            }
        }
    }

    $\boldsymbol{\theta}_{\text{pbest}} \gets \boldsymbol{\theta}, \boldsymbol{c_{\text{pbest}}} \gets \boldsymbol{c}$\;

    \For{$i \gets 1$ \textbf{to} $m$}{
        $v \gets w_{\text{ine}} * v + w_1 * \text{rand(p,n)} * (\theta_{\text{pbest}} - \theta) + w_2 * \text{rand(p,n)} * (\theta_{\text{gbest}} - \theta)$\tcp*[r]{Move particles}
        $\theta \gets \theta + v$\; 

        \For{$j \gets 1$ \textbf{to} $p$}{
            \For{$k \gets 1$ \textbf{to} $n$}{
                \If{$\boldsymbol{\theta}(j, k) < \boldsymbol{\theta}_{\mathrm{min}}(k)$}{
                    $\boldsymbol{\theta}(j, k) \gets \boldsymbol{\theta}_{\mathrm{min}}(k)$\;
                }
                \If{$\boldsymbol{\theta}(j, k) > \boldsymbol{\theta}_{\mathrm{max}}(k)$}{
                    $\boldsymbol{\theta}(j, k) \gets \boldsymbol{\theta}_{\mathrm{max}}(k)$\;
                }
                $\boldsymbol{\theta}_{\text{aux}}(k) \gets \boldsymbol{\theta}(j, k)$\;
            }
            $\boldsymbol{c}(j) \gets \Cost(\boldsymbol{\theta}_{\text{aux}}, \lambda, G, \boldsymbol{s}, f(x), \boldsymbol{y}^{(r)})$\;
            \If{$\boldsymbol{c}(j) < \boldsymbol{c}_{\mathrm{pbest}}(j)$}{
                $\boldsymbol{c}_{\mathrm{pbest}}(j) \gets \boldsymbol{c}(j)$\;
                \For{$k \gets 1$ \textbf{to} $n$}{
                    $\boldsymbol{\theta}_{\text{pbest}}(j, k) \gets \boldsymbol{\theta}(j, k)$\;
                }
            }
            \If{$\boldsymbol{c}(j) < c_{\mathrm{gbest}}$}{
                $c_{\text{gbest}} \gets \boldsymbol{c}(j)$\;
                \For{$k \gets 1$ \textbf{to} $n$}{
                    $\boldsymbol{\theta}_{\text{gbest}}(k) \gets \boldsymbol{\theta}(j, k)$\;
                }
            }
        }
        $w_{\text{ine}} \gets w_{\text{ine}} * 0.7$\; 
    }
    \Return $c_{\text{gbest}}, \boldsymbol{\theta}_{\text{gbest}}$\;
}

\end{algorithm}

\section{Conclusion} \label{conclusion}
This paper studied a novel, low-cost, and easily deployable sub-THz RoF system implementation that combines PMFs and RUs in a cascaded structure. A UL signal model of this system was developed by considering two regimes for PAs, linear and non-linear. For our analysis, we considered two types of fibers: flat fibers and frequency-selective fibers, whose effects were included in the signal model. Based on this model, an ML framework and an NLS framework were developed to estimate the propagation distance along the RoF and the TOA of signals transmitted from a UE. The CRLB was derived as a benchmark of the variance of the proposed ML estimator. 

Monte-Carlo simulations were performed to assess the performance of the proposed estimator. Our results demonstrate that good performance of the proposed estimators can be achieved under the effects of cascaded PMFs and RUs. The ML estimator achieves superior performance for a frequency-selective fiber compared to a frequency-flat fiber. Finally, our proposed algorithms show high accuracy in positioning UEs in an indoor scenario.

A future direction could be estimating how the second moment of the variance is affected by the non-linearities and including that in the estimation criterion. Another possible direction is to replace the third-order polynomial model of PA non-linearity with a more accurate one. Also, considering the antenna radiation patterns, such as a highly directional antenna, and handling the UE's orientation, enhances the research. 

\section*{Acknowledgment}
We thank Dr. Frida Strömbeck and Prof. Herbert Zirath from Chalmers University of Technology for generously sharing the measured PMF characteristics that we used in our numerical examples.

\vspace*{-\baselineskip}
\begin{appendix} \label{append}
In this appendix, we provide the detailed derivations of the CRLB shown in Section~\ref{Cramer-Rao Lower bound}. Specifically, we write the respective derivations of the CRLB for both the flat-fiber and frequency-selective fiber assumptions. Note that both CRLBs are obtained from \eqref{variance of wk}. The two types of fiber show different characteristics. Hence, we present the CRLB for each case separately. We define $\boldsymbol{\Omega}_k \triangleq \phi+r \psi_k-2 \pi f_k \tau$.

\section*{CRLB for flat fiber}
Partial derivatives in \eqref{FIM definition} can be calculated as 

\begin{subequations}
\begin{align}
         \left[\frac{\partial \boldsymbol{\mu(\theta)}}{\partial \lvert A \rvert}\right]_{k}=&G e^{j(\boldsymbol{\Omega}_k)} s_k, \tag{40a} \\
        \left[\frac{\partial \boldsymbol{\mu(\theta)}}{\partial \phi}\right]_{k}=& j G \lvert A \rvert  s_k e^{j(\boldsymbol{\Omega}_k)}, \tag{40b}\\
        \left[\frac{\partial \boldsymbol{\mu(\theta)}}{\partial \tau}\right]_{k}=&-j 2 \pi f_k G \lvert A \rvert  s_k e^{j(\boldsymbol{\Omega}_k)},\tag{40c}\\
        \left[\frac{\partial \boldsymbol{\mu(\theta)}}{\partial r}\right]_{k}=& G \lvert A \rvert e^{j(\boldsymbol{\Omega}_k)}s_k \left(j\psi_k \right).\tag{40d}
\end{align}
\end{subequations}
The derivations of CRLB are as follows. 
\begin{subequations}
\begin{align}
        \left[\boldsymbol{I}(\theta)\right]_{\lvert A \rvert,\lvert A \rvert}=& 2\Re\left\{\!\frac{G^{(2r+2)}}{(r+1)\sigma^2}\! \sum_{k=0}^{K-1}  \lvert H_k\rvert^{2r} \! \lvert s_k \rvert^2\! e^{-j(\boldsymbol{\Omega}_k)} \! e^{j(\boldsymbol{\Omega}_k)}\!\right\}, \notag \\
        & = 2\frac{G^{2} }{(r+1)\sigma^2} \sum_{k=0}^{K-1}\lvert s_k \rvert^2, \tag{41a}\\
        \left[\boldsymbol{I}(\theta)\right]_{\lvert A \rvert,\phi}=& 2\Re\left\{\frac{j \lvert A \rvert G^2}{(r+1)\sigma^2}\sum_{k=0}^{K-1} \lvert s_k \rvert^2 e^{-j(\boldsymbol{\Omega}_k)} e^{j(\boldsymbol{\Omega}_k)}\right\}, \notag \\
        = &0, \tag{41b}\\
       \left[\boldsymbol{I}(\theta)\right]_{\lvert A \rvert,\tau}=& 2\Re\left\{-\frac{j 2 \pi G^2\lvert A \rvert }{(r+1)\sigma^2}\sum_{k=0}^{K-1} f_k \lvert s_k \rvert^2 e^{-j(\boldsymbol{\Omega}_k)}e^{j(\boldsymbol{\Omega}_k)}\right\},\notag\\
       =& 0,\tag{41c}
           \end{align}
    \end{subequations}
    \begin{subequations}
     \begin{align}     
        \left[\boldsymbol{I}(\theta)\right]_{\lvert A \rvert,r}=&2\Re\left\{\frac{G^2 \lvert A \rvert}{(r+1) \sigma^2}\sum_{k=0}^{K-1} \lvert s_k\rvert^2 \left(j\psi_k  \right)\right\}, \notag\\
       =& 0, \tag{41d}\\
        \left[\boldsymbol{I}(\theta)\right]_{\phi,\phi}=&2\Re\left\{\frac{-j^2 \lvert A \rvert^2 G^2 }{(r+1)\sigma^2} \sum_{k=0}^{K-1} \lvert s_k \rvert^2 e^{-j(\boldsymbol{\Omega}_k)} e^{j(\boldsymbol{\Omega}_k)}  \right\},\notag\\
        =& 2\frac{G^2 \lvert A \rvert^2}{(r+1)\sigma^2} \sum_{k=0}^{K-1}  \lvert s_k \rvert^2, \tag{41e}\\
         \left[\boldsymbol{I}(\theta)\right]_{\phi,\tau}=&2\Re\left\{\frac{j^2 2 \pi \lvert A \rvert^2 G^2 }{(r+1)\sigma^2} \sum_{k=0}^{K-1} f_k  \lvert s_k\rvert^2 e^{-j(\boldsymbol{\Omega}_k)} e^{j(\boldsymbol{\Omega}_k)}\right\}, \notag \\ =& -\frac{ 4\pi G^2 \lvert A \rvert^2}{(r+1)\sigma^2} \sum_{k=0}^{K-1} f_k   \lvert s_k \rvert^2, \tag{41f}
    \end{align}
    \end{subequations}
    \begin{subequations}
     \begin{align}     
    \left[\boldsymbol{I}(\theta)\right]_{\phi,r}=&2\Re\left\{\frac{-j \lvert A \rvert^2 G^2}{(r+1)\sigma^2} \sum_{k=0}^{K-1} \lvert s_k \rvert^2 \left(j\psi_k \right)\right\},\notag\\
        =& 2\frac{G^{2} \lvert A \rvert^2 }{(r+1)\sigma^2} \sum_{k=0}^{K-1} \lvert s_k \rvert^2 \psi_k , \tag{41g}\\
         \left[\boldsymbol{I}(\theta)\right]_{\tau,\tau}=& 2\Re\left\{\frac{-j^2 4 \pi^2 \lvert A \rvert^2 G^2}{(r+1) \sigma^2}\sum_{k=0}^{K-1} f_k^2 \lvert s_k \rvert^2 e^{-j(\boldsymbol{\Omega}_k)} e^{j(\boldsymbol{\Omega}_k)} \right\},\notag\\
       =&\frac{8\pi^2 G^2 \lvert A \rvert^2 }{(r+1)\sigma^2} \sum_{k=0}^{K-1} f_k^2 \lvert s_k\rvert^2,\tag{41h}\\
     \left[\boldsymbol{I}(\theta)\right]_{r,r}=& K\left\{\frac{\sigma^2}{(r+1)\sigma^2} \frac{\sigma^2}{(r+1)\sigma^2}\right\}\notag\\
       &+2\Re\left\{\frac{ \lvert A \rvert^2 G^2}{(r+1)\sigma^2} \sum_{k=0}^{K-1} \lvert s_k \rvert^2 \left(-j\psi_k\right) \left(j\psi_k\right) \right\}, \notag\\
       =&\frac{K}{(r+1)^2}+2\frac{G^{2} \lvert A\rvert^2}{(r+1) \sigma^2} \sum_{k=0}^{K-1} \lvert s_k\rvert^2 \psi_k^2, \tag{41i}\\
       \left[\boldsymbol{I}(\theta)\right]_{\tau,r}=& 2\Re\left\{\frac{j 2 \pi \lvert A \rvert^2 G^2}{(r+1)\sigma^2}\sum_{k=0}^{K-1} f_k \lvert s_k \rvert^2 \left(j\psi_k \right)\right\},\notag\\
       =& -\frac{4\pi G^2 \lvert A \rvert^2}{(r+1)\sigma^2} \sum_{k=0}^{K-1}\lvert s_k \rvert^2 f_k \psi_k. \tag{41j}
\end{align}
\end{subequations}

\section*{CRLB for frequency-selective fiber}
One can derive the first-order derivatives in \eqref{mean vector selective} as 
\begin{subequations}
\begin{align}
     \left[\frac{\partial \boldsymbol{\mu(\theta)}}{\partial \lvert A \rvert}\right]_{k}&=G^{(r+1)} \lvert H_k \rvert^r e^{j(\phi+r \psi_k-2 \pi f_k \tau)} s_k, \tag{42a} \\
    \left[\frac{\partial \boldsymbol{\mu(\theta)}}{\partial \phi}\right]_{k}& = j G^{(r+1)} \lvert A \rvert \lvert H_k \rvert^r s_k e^{j(\phi+r \psi_k-2 \pi f_k \tau)}, \tag{42b}\\
        \left[\frac{\partial \boldsymbol{\mu(\theta)}}{\partial \tau}\right]_{k}\!&=\!-j 2 \pi f_k G^{(r+1)} \!\lvert A \rvert \lvert H_k \rvert^r s_k e^{j(\phi+r \psi_k-2 \pi f_k \tau)},\tag{42c}\\
        \left[\frac{\partial \boldsymbol{\mu(\theta)}}{\partial r}\right]_{k}\notag &= \! G^{(r+1)} \lvert A \rvert \lvert H_k \rvert^r \! s_k e^{j(\phi+r \psi_k-2 \pi f_k \tau)} \\
       &\times \left(\!\ln\!{\sqrt{b_k}} \! +\!j\psi_k\!\right)\!.\tag{42d}
       \end{align}
\end{subequations}
The derivations of CRLB as as follows. 
\begin{subequations}
\begin{align}
        \left[\boldsymbol{I}(\theta)\right]_{\lvert A \rvert,\lvert A \rvert}=& 2\Re\left\{\!G^2\!\sum_{k=0}^{K-1}\!\frac{(b_k\!-\!1)b_k^r\lvert s_k \rvert^2 \!e^{-j(\boldsymbol{\Omega}_k)} e^{j(\boldsymbol{\Omega}_k)}}{\sigma^2 (b_k^{r+1}-1)}\!\right\}, \notag \\
        & = 2G^2\sum_{k=0}^{K-1}\frac{(b_k-1)b_k^r\lvert s_k \rvert^2}{\sigma^2 (b_k^{r+1}-1)},\tag{43a} \\
        \left[\boldsymbol{I}(\theta)\right]_{\lvert A \rvert,\phi}=& 2\Re\left\{j \lvert A \rvert G^2\sum_{k=0}^{K-1} \frac{(b_k-1) b_k^r \lvert s_k \rvert^2 e^{-j(\boldsymbol{\Omega}_k)} e^{j(\boldsymbol{\Omega}_k)}}{\sigma^2(b_k^{r+1}-1)}\right\}, \notag \\
        = &0,  \tag{43b}\\
        \left[\boldsymbol{I}(\theta)\right]_{\lvert A \rvert,\tau}=& 2\Re\left\{\!-j 2 \pi G^2\lvert A \rvert\! \sum_{k=0}^{K-1}\!\frac{(b_k-1) f_k b_k^r \lvert s_k \rvert^2 e^{-j(\boldsymbol{\Omega}_k)}e^{j(\boldsymbol{\Omega}_k)}}{\sigma^2 (b_k^{r+1}-1)}\!\right\},\notag\\
       =& 0, \tag{43c}\\
        \left[\boldsymbol{I}(\theta)\right]_{\lvert A \rvert,r}=&2\Re\left\{G^2\lvert A \rvert \sum_{k=0}^{K-1}\frac{(b_k-1) b_k^{r}\lvert s_k\rvert^2  \left(\ln{\lvert G H_k \rvert} +j\psi_k  \right)}{\sigma^2(b_k^{r+1}-1)}\right\}, \notag\\
       =& 2 G^{2} \lvert A \rvert \sum_{k=0}^{K-1} \frac{(b_k-1) b_k^{r} \lvert s_k\rvert^2 \ln{\sqrt{b_k}}}{\sigma^2(b_k^{r+1}-1)}, \tag{43d}\\
        \left[\boldsymbol{I}(\theta)\right]_{\phi,\phi}=&2\Re\left\{\lvert A \rvert^2 G^{2} \sum_{k=0}^{K-1} \frac{(b_k-1)b_k^{r} \lvert s_k \rvert^2 e^{-j(\boldsymbol{\Omega}_k)} e^{j(\boldsymbol{\Omega}_k)}}{\sigma^2(b_k^{r+1}-1)} \right\},\notag\\
        =& 2 G^{2} \lvert A \rvert^2\sum_{k=0}^{K-1} \frac{(b_k-1)b_k^{r}\lvert s_k \rvert^2}{\sigma^2(b_k^{r+1}-1)},  \tag{43e}\\
        \left[\boldsymbol{I}(\theta)\right]_{\phi,\tau}=&2\Re\left\{\!- 2 \pi \lvert A \rvert^2 G^{2}\!\sum_{k=0}^{K-1}\!\frac{ (b_k-1) f_k b_k^{r} \lvert s_k \rvert^2 e^{-j(\boldsymbol{\Omega}_k)} e^{j(\boldsymbol{\Omega}_k)}}{\sigma^2(b_k^{r+1}-1)}\!\right\}, \notag \\
        =& -4 \pi G^{2} \lvert A \rvert^2\sum_{k=0}^{K-1}\frac{(b_k-1) f_k b_k^{r} \lvert s_k \rvert^2}{\sigma^2(b_k^{r+1}-1)}, \tag{43f}\\
         \left[\boldsymbol{I}(\theta)\right]_{\phi,r}=&2\Re\left\{\!-j \lvert A \rvert^2 G^{2}\!\sum_{k=0}^{K-1}\!\frac{(b_k-1)b_k^{r}\lvert s_k \rvert^2 \left(\ln{\lvert G H_k \rvert} +j\psi_k \right)}{\sigma^2(b_k^{r+1}-1)} \!\right\},\notag\\
        =& 2 G^{2} \lvert A \rvert^2 \sum_{k=0}^{K-1} \frac{(b_k-1) b_k^{r} \lvert s_k \rvert^2 \psi_k}{\sigma^2(b_k^{r+1}-1)}, \tag{43g} \\
         \left[\boldsymbol{I}(\theta)\right]_{\tau,\tau}=&2\Re\left\{ \!4 \pi^2 \lvert A \rvert^2 G^{2}\!\sum_{k=0}^{K-1}\!\frac{(b_k-1)f_k^2 b_k^{r} \lvert s_k \rvert^2 e^{-j(\boldsymbol{\Omega}_k)} e^{j(\boldsymbol{\Omega}_k)}}{\sigma^2(b_k^{r+1}-1)} \!\right\},\notag\\
       =& 8\pi^2 G^{2} \lvert A \rvert^2 \sum_{k=0}^{K-1}\frac{ (b_k-1) f_k^2 b_k^{r} \lvert s_k\rvert^2}{\sigma^2(b_k^{r+1}-1)}, \tag{43h}
    \end{align}
\end{subequations}

\begin{figure*}[ht]
\begin{align}
       \left[\boldsymbol{I}(\theta)\right]_{r,r}=& \mathrm{tr}\left\{\boldsymbol{C}(\boldsymbol{\theta})^{-1}\frac{\partial \boldsymbol{C}(\boldsymbol{\theta})}{\partial r} \boldsymbol{C}(\boldsymbol{\theta})^{-1} \frac{\partial \boldsymbol{C}(\boldsymbol{\theta})}{\partial r}\right\}+2\Re\left\{\lvert A \rvert^2 G^{2}\sum_{k=0}^{K-1} \frac{(b_k-1)\lvert s_k \rvert^2 b_k^{r} \left(\ln{\lvert G H_k \rvert} -j\psi_k\right) \left( \ln{\lvert G H_k \rvert}+j\psi_k\right)}{\sigma^2(b_k^{r+1}-1)} \right\}, \notag\\
       =&\sum_{k=0}^{K-1}\left(\frac{b_k^r\ln{b_k}}{1-b_k^{r+1}}\right)^2+2 G^{2} \lvert A\rvert^2 \sum_{k=0}^{K-1} \frac{\lvert s_k\rvert^2 (b_k-1) b_k^{r} \left( (\ln{\sqrt{b_k}})^2+\psi_k^2 \right)}{\sigma^2(b_k^{r+1}-1)}, \tag{43i}\\
            \left[\boldsymbol{I}(\theta)\right]_{\tau,r}=& 2\Re\left\{j 2 \pi \lvert A \rvert^2 G^{2}\sum_{k=0}^{K-1} \frac{(b_k-1) f_k b_k^{r} \lvert s_k \rvert^2 \left(\ln{\lvert G H_k \rvert} +j\psi_k \right)}{\sigma^2(b_k^{r+1}-1)}\right\},\notag\\
       =& - 4\pi G^{2}\lvert A \rvert^2\sum_{k=0}^{K-1}\frac{(b_k-1) b_k^{r} \lvert s_k \rvert^2 f_k \psi_k}{\sigma^2(b_k^{r+1}-1)}.\tag{43j}
\end{align}
\end{figure*}
\end{appendix}

\vspace*{130pt}
\bibliographystyle{IEEEtran}
\bibliography{reference}

\begin{IEEEbiography}[{\includegraphics[width=1in,height=1.25in,clip,keepaspectratio]{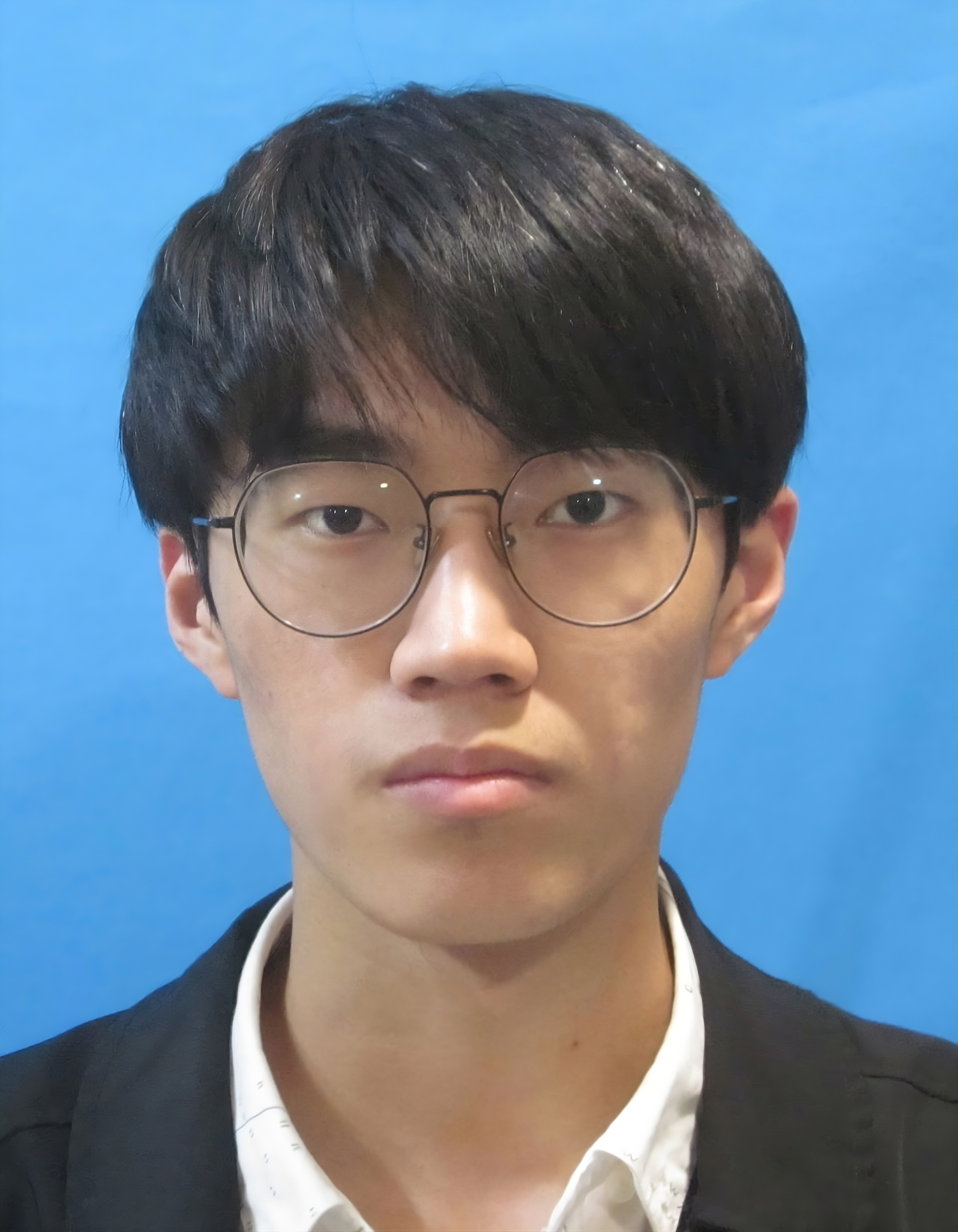}}]{Dexin Kong} (Graduate Student Member, IEEE) received his B.Eng. degree in communication engineering from Nanjing University of Posts and Telecommunications (NJUPT), Nanjing, China, in 2021, and M.Sc. degree in communication systems from Lund University, Lund, Sweden, in 2023. He is currently pursuing a Ph.D. degree with the Department of Electrical Engineering at Link\"oping University, Link\"oping, Sweden. His research interests include wireless communications, sub-terahertz communications, and radio-over-fiber technologies.     
    
\end{IEEEbiography}

\begin{IEEEbiography}[{\includegraphics[width=1in,height=1.25in,clip,keepaspectratio]{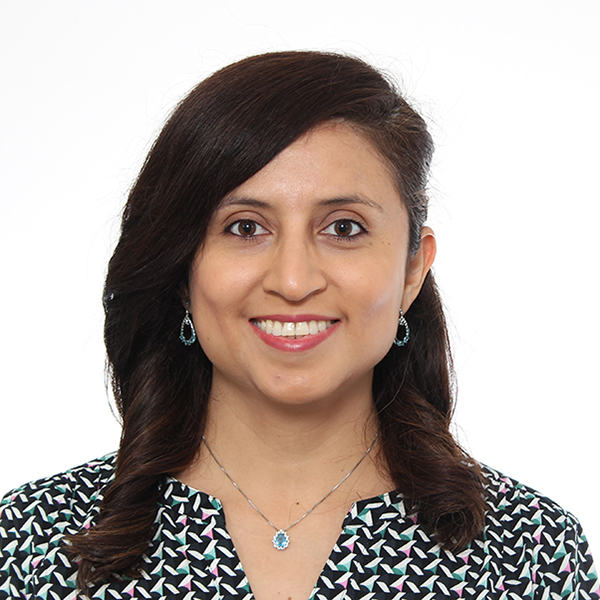}}]{Diana Pamela Moya Osorio} (M'16, SM’23) is currently Associate Professor at the Communication Systems Division, Department of Electrical Engineering, Linköping University, Sweden, and an ELLIIT recruited faculty. Previously, she was Senior Research Fellow and Adjunct Professor at the Centre for Wireless Communications, University of Oulu, Finland. She received the B.Sc. degree in electronics and telecommunications engineering from the Armed Forces University, Ecuador, in 2008, and the M.Sc. and D.Sc. degrees in electrical engineering with emphasis on telecommunications and telematics from the University of Campinas, Brazil, in 2011 and 2015, respectively. From 2015 to 2022, she was an Assistant Professor with the Department of Electrical Engineering, Federal University of São Carlos, Brazil. From 2020 to 2023, she was also a Postdoctoral Researcher for the Academy of Finland. She has served as TPC and reviewer for several journals and conferences. Currently, she is Associate Editor of IEEE Communication Letters and IEEE Transactions on Information Forensics \& Security. She also serves as working group leader for Trustworthy 6G at the Cost Action 6G-PHYSEC. Her research interests include wireless communications and sensing in general, signal processing for wireless communications and radar systems, physical layer security and trustworthy systems, and integrated sensing and communications.
\end{IEEEbiography}

\begin{IEEEbiography}[{\includegraphics[width=1in,height=1.25in,clip,keepaspectratio]{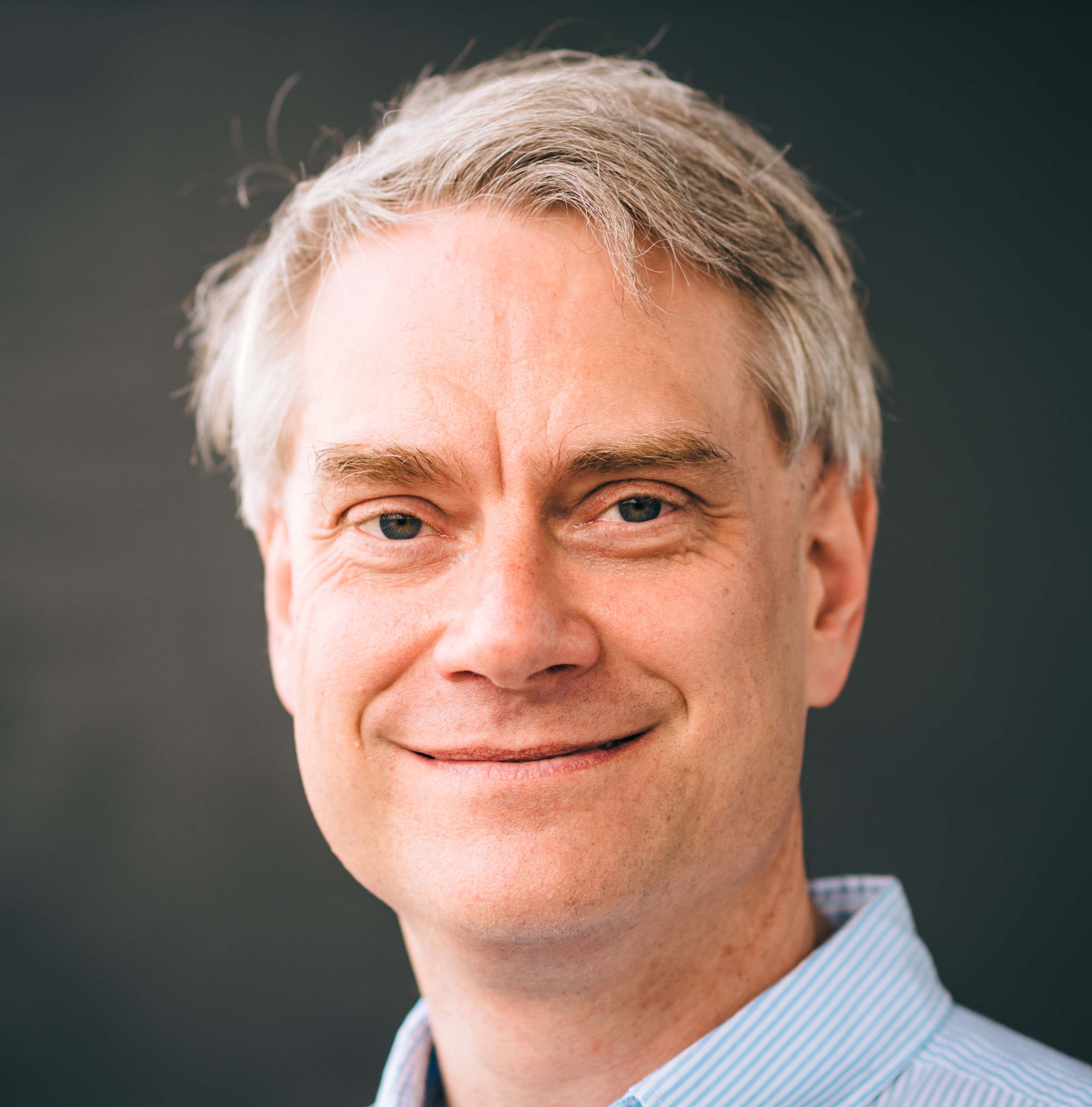}}]
 {Erik G. Larsson} (Fellow, IEEE) received the Ph.D. degree from Uppsala University, Uppsala, Sweden, in 2002.  He is currently Professor of Communication Systems at Link\"oping University (LiU) in Link\"oping, Sweden. He was with the KTH Royal Institute of Technology in Stockholm, Sweden, the George Washington University, USA, the University of Florida, USA, and Ericsson Research, Sweden.  His main professional interests are within wireless communications, signal processing, network science, and decentralized machine learning. He co-authored \emph{Space-Time Block Coding for  Wireless Communications} (Cambridge University Press, 2003) and \emph{Fundamentals of Massive MIMO} (Cambridge University Press, 2016). He served as chair of the IEEE Signal Processing Society SPCOM technical committee (2015--2016), chair of the \emph{IEEE Wireless  Communications Letters} steering committee (2014--2015), member of the  \emph{IEEE Transactions on Wireless Communications} steering committee (2019-2022), General and Technical Chair of the Asilomar SSC conference (2015, 2012), technical co-chair of the IEEE Communication Theory Workshop (2019), and member of the  IEEE Signal Processing Society Awards Board (2017--2019). He was Associate Editor for, among others, the \emph{IEEE Transactions on Communications} (2010-2014), the \emph{IEEE Transactions on Signal Processing} (2006-2010), and the \emph{IEEE Signal  Processing Magazine} (2018-2022). He received the IEEE Signal Processing Magazine Best Column Award twice, in 2012 and 2014, the IEEE ComSoc Stephen O. Rice Prize in Communications Theory in 2015, the IEEE ComSoc Leonard G. Abraham Prize in 2017, the IEEE ComSoc Best Tutorial Paper Award in 2018, the IEEE ComSoc Fred W. Ellersick Prize in 2019, and the IEEE SPS Donald G. Fink Overview Paper Award in 2023. He is a Fellow of EURASIP, a member of the Swedish Royal Academy of Sciences (KVA), and Highly Cited according to ISI Web of Science.    
\end{IEEEbiography}

\end{document}